\documentclass{article}

\usepackage[english]{babel}

\usepackage[letterpaper,top=2cm,bottom=2cm,left=2cm,right=2cm,marginparwidth=1.75cm]{geometry}

\usepackage{amsmath}
\usepackage{graphicx}

\usepackage{caption}
\usepackage{subcaption}
\usepackage[per-mode=symbol]{siunitx} 
\DeclareSIUnit{\eaBohr}{$ea_0$}
\usepackage{makecell}
\usepackage{braket}
\usepackage{multirow}
\usepackage{csquotes}
\usepackage{orcidlink}
\usepackage{hyperref}

\let\vec\mathbf

\usepackage[
  backend=biber,
  style=phys,
  articletitle=true,
  doi=false,
  url=false,
  sorting=none
]{biblatex}

\DeclareFieldFormat{note}{
  \iffieldundef{url}
    {#1}
    {\href{\thefield{url}}{#1}}
}
\DeclareFieldFormat{type}{}

\AtBeginBibliography{
  \renewcommand*{\mkbibsuperscript}[1]{}
}
\DeclareFieldFormat{labelnumberwidth}{\mkbibbrackets{#1}}
\usepackage{fancyhdr}
\title{$^{88}$Sr Reference Data}
\author{S.\,Pucher, S.\,L.\,Kristensen, R.\,M.\,Kroeze}

\begin{document}
\thispagestyle{empty}

\begin{center}
  {\LARGE\bf $^{88}$Sr Reference Data}

  \vspace{3em}

  Sebastian Pucher\orcidlink{0000-0001-5559-0114},\textsuperscript{1,2$\dagger$}\;
  Sofus Laguna Kristensen\orcidlink{0000-0002-2463-9883},\textsuperscript{1,2$\ddagger$}\;
  Ronen M. Kroeze\orcidlink{0009-0007-9976-5124}\textsuperscript{2,3*}

  \vspace{1em}

  \small\textsuperscript{1}Max-Planck-Institut für Quantenoptik, 85748 Garching, Germany\\
  \small\textsuperscript{2}Munich Center for Quantum Science and Technology, 80799 M{\"u}nchen, Germany\\
  \small\textsuperscript{3}Fakult{\"a}t für Physik, Ludwig-Maximilians-Universit{\"a}t, 80799 M{\"u}nchen, Germany

  \vspace{0.5em}

  (Dated: \today)

  \vspace{1em}

  \begin{minipage}{0.85\textwidth}
Strontium-88 is a versatile atomic species often used in quantum optics, precision metrology, and quantum computing. Consolidated atomic data is essential for the planning, execution, and evaluation of experiments. In this reference, we present physical and optical properties of neutral $^{88}$Sr relevant to these applications. Here we focus on experimental results and supplement these with theoretical values. We present equations to convert values and derive important parameters. Tabulated results include key parameters for commonly used transitions in $^{88}$Sr ($^1\mathrm{S}_0 \rightarrow \, ^1\mathrm{P}_1$, $^1\mathrm{S}_0 \rightarrow \, ^3\mathrm{P}_{0,1,2}$, and $^3\mathrm{P}_{0,1,2} \rightarrow \, ^3\mathrm{S}_1$). This dataset serves as an up-to-date reference for studies involving bosonic $^{88}$Sr.
  \end{minipage}

\end{center}

\begin{figure}[b]
  \noindent\rule{0.35\textwidth}{0.4pt}

  \footnotesize
  \noindent
  \textsuperscript{$\dagger$} Email: sebastian.pucher@protonmail.com, Present address: \href{https://ror.org/05vxfdm89}{Atom Computing, Inc.}, 2500 55th St, Boulder, Colorado 80301, USA\\
  \textsuperscript{$\ddagger$} Email: sofus.kristensen@mpq.mpg.de, Present address: \href{https://ror.org/05vxfdm89}{Atom Computing Denmark ApS}, Ronnegade 4, 2100 Copenhagen, Denmark\\
  \textsuperscript{*} Email: r.kroeze@lmu.de
\end{figure}

\section{Introduction}

In this reference, we present the physical and optical properties of strontium-88 ($^{88}$Sr) relevant to various quantum optics, laser cooling, and precision measurement experiments. This document is inspired by the widely used references on alkali atoms by Daniel A.\ Steck~\cite{Steck2024SodiumDLine, Steck2024Rb85DLine, Steck2024Rb87DLine, Steck2024CesiumDLine}. Our goal is to provide experimentalists and theorists with consolidated data and calculated parameters that are important for understanding the interaction of light with strontium atoms.

We consider the relevant electronic level structure of $^{88}$Sr, in the context of laser cooling, optical trapping, coherent manipulation, and optical clock operation. In particular, we summarize the following optical transitions:
\begin{itemize}
    \item The broad, dipole-allowed $5\mathrm{s}^2\;{}^1\mathrm{S}_0 \rightarrow 5\mathrm{s}\,5\mathrm{p}\;{}^1\mathrm{P}_1$ transition at a wavelength $\lambda$ of about \SI{461}{\nano\meter}, which is used for first-stage laser cooling and Zeeman slowing (blue MOT transition), but also commonly for atomic imaging schemes.
    \item The narrow spin-forbidden intercombination $5\mathrm{s}^2\;{}^1\mathrm{S}_0 \rightarrow 5\mathrm{s}\,5\mathrm{p}\;{}^3\mathrm{P}_1$ transition at $\lambda\approx\SI{689}{\nano\meter}$, used for second-stage cooling to sub-microkelvin temperatures (red MOT transition).
    \item The additional dipole-allowed transitions at $\lambda\approx\SI{679}{\nano\meter}$, \SI{688}{\nano\meter}, and \SI{707}{\nano\meter} are used to excite atoms into the $5\mathrm{s}\,6\mathrm{s}\;{}^3\mathrm{S}_1$ state, which is primarily used to manipulate and repump population of the long-lived metastable $5\mathrm{s}\,5\mathrm{p}\;{}^3\mathrm{P}_0$ and $5\mathrm{s}\,5\mathrm{p}\;{}^3\mathrm{P}_2$ states back into the cooling cycle.
    \item The ultra-narrow $5\mathrm{s}^2\;{}^1\mathrm{S}_0 \rightarrow 5\mathrm{s}\,5\mathrm{p}\;{}^3\mathrm{P}_0$ transition at $\lambda\approx\SI{698}{\nano\meter}$, used in optical lattice clocks for precision timekeeping and as an optical qubit for quantum information applications.
    \item The ultra-narrow $5\mathrm{s}^2\;{}^1\mathrm{S}_0 \rightarrow 5\mathrm{s}\,5\mathrm{p}\;{}^3\mathrm{P}_2$ transition at $\lambda\approx\SI{671}{\nano\meter}$, relevant for state preparation, manipulation, and quantum simulations.
\end{itemize}

Measured values are provided with their corresponding references to original experimental sources, while calculated quantities are derived using established physical models, with citations to more detailed theoretical discussions. The data provided here are not guaranteed to be complete or error-free. For parameters critical to your research, you should consult the primary literature.

This document is intended as a community resource, following the example of Steck’s \textit{Alkali D Line Data} series, available at \url{http://steck.us/alkalidata}. We hope that this summary will serve experimental and theoretical researchers working with strontium, whether for atomic physics, quantum simulation, or metrology. Comments, corrections, and suggestions for improving this document are welcomed and encouraged.

\section{Statistical Methods}

This section outlines the statistical methods used to combine and report the numerical values derived from multiple experimental or theoretical sources. If more than one value is cited for quantities such as the transition frequency, the lifetime of a state, the isotope shift, or the scattering length, the quoted value is a weighted mean of the values from the references, calculated as
\begin{equation}
  \mu = \frac{\sum_{j=1}^n x_j/\sigma_j^2}{\sum_{j=1}^n 1/\sigma_j^2}~,
\end{equation}
where \( x_j \) are the measurement values with uncertainties \( \sigma_j \), and \((j=1,\dots,n)\) indexes the individual references. Here, the weights are the inverse squares of the uncertainties. The corresponding standard uncertainty in the weighted mean is
\begin{equation}
  \sigma_\mu = \sqrt{ \frac{1}{\sum_{j=1}^n 1/\sigma_j^2} }~.
\end{equation}
If the measurements are not statistically consistent (i.e., the scatter exceeds what is expected from the reported uncertainties), we scale the uncertainty in the mean by the Birge ratio~\cite{Birge1932}, defined as
\begin{equation}
  R_{\mathrm{B}} = \sqrt{ \frac{1}{n-1} \sum_{j=1}^n \frac{(x_j - \mu)^2}{\sigma_j^2} } = \sqrt{ \frac{\chi^2}{n-1} }~,
\end{equation}
where \( \chi^2 \) is the chi-squared value and \( n \) is the number of data points. The final uncertainty in the mean is then
\begin{equation}
  \sigma_\mu^{\mathrm{adj}} = R_{\mathrm{B}} \cdot \sigma_\mu~.
\end{equation}
If \( R_{\mathrm{B}} > 1 \) for one dataset, we show $R_{\mathrm{B}}$ next to the reported weighted mean to indicate that the measured values are statistically inconsistent, and we adjusted the uncertainty with the Birge ratio. We note that alternative techniques for assessing the uncertainty of a weighted mean are actively being investigated and may be employed in future updates~\cite{Trassinelli2024}.

When an experimental reference reports both a statistical uncertainty \(\sigma_{\rm stat}\) and a systematic uncertainty \(\sigma_{\rm sys}\) for a single measurement, the uncertainties are assumed to be uncorrelated. Hence, we combine the two contributions in quadrature to give the total one‐sigma uncertainty
\begin{equation}
  \sigma = \sqrt{\sigma_{\rm stat}^2 + \sigma_{\rm sys}^2}~.
\end{equation}
In cases where only asymmetric limits are provided, e.g. ${x_0}^{+\Delta x_+\!}_{-\Delta x_-\!}$, we obtain a single symmetric uncertainty by taking the larger deviation
\begin{equation}
  \sigma = \max\bigl(\Delta x_+,\,\Delta x_-\bigr)~.
\end{equation}
Reporting $x_0(\sigma) = x_0 \pm \sigma$ then guarantees coverage of both published bounds. Although this approach overestimates the error when the true distribution is asymmetric, it provides a straightforward, conservative uncertainty for tables and figures.

\section{Strontium Properties}

Strontium is a soft, silver-white alkaline-earth metal. Freshly cut, it displays a bright silvery luster that quickly turns yellow as surface oxide forms. Below \SI{380}{\celsius}, it does not absorb nitrogen, but it should be stored in, e.g., mineral oil or an argon atmosphere to prevent oxidation~\cite{crc97}. Strontium reacts vigorously with water~\cite{crc97}. Contaminated parts and tools can be cleaned with water to remove unwanted strontium, but this should be done with caution due to the rapid reaction of the metal with water. In finely divided form, the metal can spontaneously ignite in air~\cite{crc97}. 

Strontium has 38 electrons, two of which occupy the outermost shell. Natural strontium consists of the four stable isotopes $^{84}\mathrm{Sr}$, $^{86}\mathrm{Sr}$, $^{87}\mathrm{Sr}$, and $^{88}\mathrm{Sr}$, alongside more than thirty known unstable isotopes. This review focuses on the most abundant stable isotope $^{88}$Sr.

Table~\ref{tab:fundamental_constants} lists selected fundamental physical constants from the 2022 CODATA recommended values that we use in this reference~\cite{Mohr2025}. In Table~\ref{tab:sr_physical_properties}, key physical properties of the isotope $^{88}\mathrm{Sr}$ are summarized. For this isotope, the nuclear spin vanishes resulting in bosonic particle statistics. The mass is the recommended value from the 2020 Atomic Mass Evaluation (AME2020)~\cite{Wang2021,Kondev2021}. The value is based on measurements of cyclotron frequency ratios of ion pairs simultaneously confined in a Penning trap~\cite{Rana2012}.

\begin{figure}[h]
    \centering
    \includegraphics[width=0.8\textwidth]{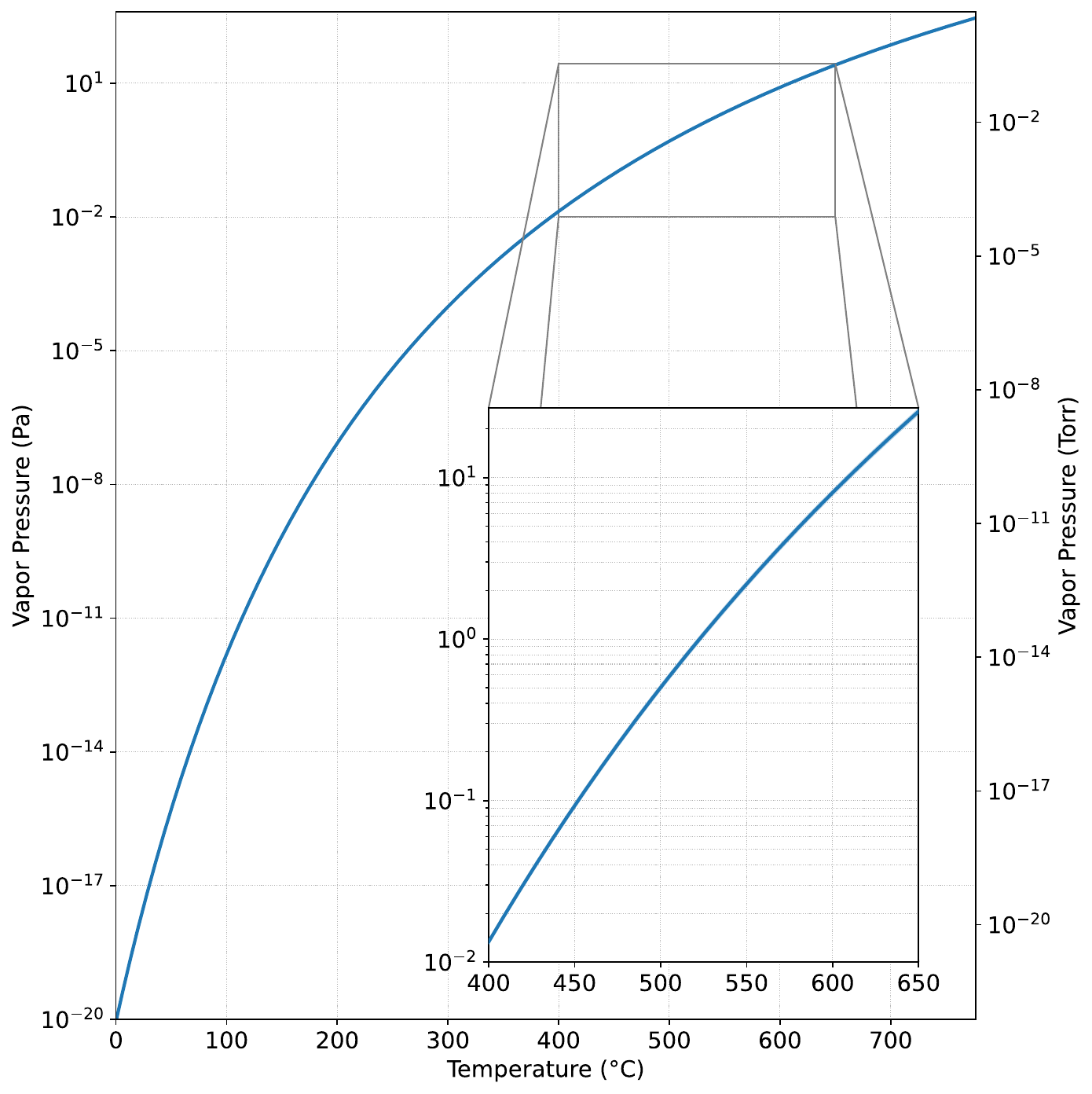}
    \caption{Vapor pressure of strontium as a function of temperature, calculated using Eq.~\ref{eq:vapor_pressure}.}
    \label{fig:vapor_pressure}
\end{figure}

The relative natural abundance, the specific heat capacity, and the molar heat capacity are taken from Ref.~\cite{crc97}. The density and molar volume at various temperatures can be found in Ref.~\cite{Arblaster2018}. The vapor pressure at a temperature of $T=\SI{25}{\celsius}$, as well as the temperature-dependent vapor pressure curve shown in Figure~\ref{fig:vapor_pressure}, are calculated using the semi-empirical equation describing solid strontium provided by~\cite{alcock1984vapor}
\begin{equation}
    \log_{10}(P) = 9.226 + \log_{10}(101325) - \frac{8572}{T} - 1.1926 \log_{10} \left(T\right)~,
    \label{eq:vapor_pressure}
\end{equation}
where $P$ is the vapor pressure in pascals and $T$ is the temperature in Kelvin. This equation is valid from $T=\SI{298}{\kelvin}$ up to the melting point of strontium, reported as $T_{\rm M}=\SI{1050}{\kelvin}$ (\SI{777}{\celsius})~\cite{crc97}. In this range, it reproduces the experimental vapor pressure with an accuracy better than~$\pm5\%$~\cite{alcock1984vapor}. Note that the density, the molar volume, the melting and boiling points, the heat capacities, as well as the vapor pressure are given for the naturally occurring form of Sr. Using its natural isotopic abundance, the corresponding values for $^{88}\mathrm{Sr}$ can be inferred.

The \(s\)-wave scattering length $a$ characterizes low-energy elastic collisions between atoms and plays a central role in ultracold gas experiments~\cite{Braaten2006}. Its magnitude governs the rate of thermalization during evaporative cooling. The sign of the scattering lengths determines whether the interactions are repulsive and a Bose–Einstein condensate (BEC) is stable (for $a > 0$) or the interactions are attractive and the BEC is prone to collapse (for $a < 0$)~\cite{Chin10}. In alkali atoms, $a$ can be magnetically tuned via Feshbach resonances to nearly arbitrary values $(-\infty, +\infty)$~\cite{Chin10}. In contrast, alkaline-earth atoms such as Sr have a $J=0$ ground state, making them largely insensitive to magnetic tuning. For these systems, the scattering length is given by the binding energy of the highest vibrational state of the ground-state molecular potential, which shifts slightly with isotopic mass~\cite{Stellmer2013}. However, Feshbach resonances can exist in a cold-atom ensemble of mixtures of the ground state $^1\mathrm{S}_0$ and metastable state $^3\mathrm{P}_2$, as shown in Yb~\cite{Takasu2017}. The isotope considered in this reference, $^{88}$Sr, has a small scattering length (see Tab.~\ref{tab:sr_scattering_lengths}), resulting in a low elastic collision rate. This makes evaporative cooling practically impossible, but is very beneficial for precision measurements~\cite{Stellmer2013}.

For potentials that decay faster than \(1/r^3\) as \(r \to \infty\), the scattering length is defined as the following low‐energy limit
\begin{equation}
    \lim_{k \to 0} k \cot \delta(k) \;=\; -\frac{1}{a}\,,
\end{equation}
where \(k\) is the wave number and \(\delta(k)\) is the phase shift of the outgoing spherical wave. Moreover, the elastic cross section \(\sigma_e\), at sufficiently low energies, is related to the scattering length via
\begin{equation}
\lim_{k \to 0} \sigma_e \;=\; 4\pi a^2 \,.
\end{equation}

\section{Level Structure}

\begin{figure}[h!]
  \centering
  \includegraphics[width=1.0\textwidth]{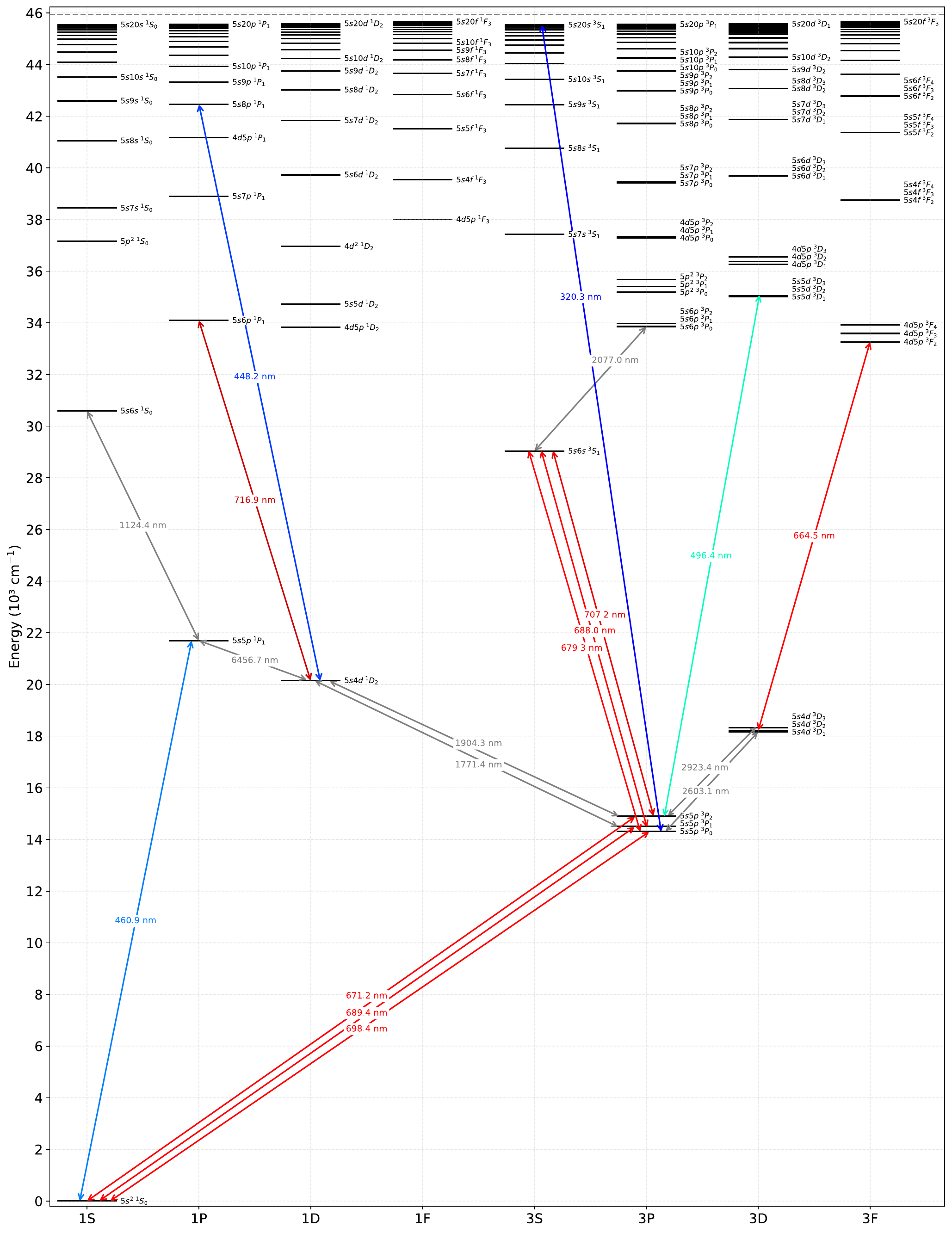}
  \caption{Neutral Sr\,I energy level diagram with selected optical transitions. Level energies and transition wavelengths are taken from the NIST Atomic Spectra Database~\cite{NIST_ASD}. The dashed gray line shows the ionization energy.}
  \label{fig:SrI-levels}
\end{figure}

As an alkaline-earth atom, Sr has two valence electrons, giving rise to a level structure of singlet and triplet terms, analogous to helium. Figure~\ref{fig:Sr_I_energy_levels_reference} shows the energy levels and transitions discussed in this reference, while Figure~\ref{fig:SrI-levels} places them within the complete Sr I level scheme. Similar level schemes can be found, e.g., in Refs.~\cite{Samland2019, Wallner2020}. The electronic configuration features both broad electric-dipole-allowed transitions with linewidths on the order of megahertz and narrow intercombination lines between singlet and triplet manifolds, with natural linewidths spanning millihertz to kilohertz.

The atomic-level notation generally consists of two parts: the electronic configuration and the term symbol \({}^{2S+1}L_J\). For the configuration, a sequence such as \(n_1l_1\,n_2l_2\dots\) indicates which orbitals are occupied by the valence electrons. For example, \(5\mathrm{s}5\mathrm{p}\) means that one electron is in the \(5\mathrm{s}\) shell and one in the \(5\mathrm{p}\) shell. In the term symbol, the superscript \(2S+1\) is the spin multiplicity, where \(S\) is the total spin quantum number (\(2S+1=1\) for singlet, \(2S+1=3\) for triplet). The letter \(L\) encodes the total orbital angular momentum (S for \(L=0\), P for \(L=1\), D for \(L=2\), ...). The subscript \(J\) is the quantum number corresponding to the total electronic angular momentum. Putting these together, a state labeled  
\begin{equation}
    n_1l_1\,n_2l_2\,\;{}^{2S+1}L_J
\end{equation}
specifies both which orbitals the valence electrons occupy and the combined spin and orbital angular momentum quantum numbers of the entire atom. In practice, when the configuration \(n_1l_1\,n_2l_2\) is unambiguous from context, we omit the explicit \(n_1l_1\,n_2l_2\) prefix and simply write the term symbol \({}^{2S+1}L_J\) for better readability. 

\subsection{Energy Level Splittings}

In $^{88}\mathrm{Sr}$, each electronic term is split by spin-orbit (fine‐structure) interactions into levels of different $J$. The total electronic angular momentum \(\mathbf{J}\) arises from coupling the orbital (\(\mathbf{L}\)) and spin (\(\mathbf{S}\)) angular momenta of the two valence electrons
\begin{equation}
  \mathbf{J} = \mathbf{L} + \mathbf{S}\,.
\end{equation}
Here, the magnitude of $\mathbf{J}$ is given by $\sqrt{J(J+1)}\,\hbar$ with possible integer values of \(J\) in the range
\begin{equation}
  |L - S| \le J \le L + S\,.  
\end{equation}
Accordingly, the \(5\mathrm{s}^2\,^1\mathrm{S}_0\) ground state of \(^{88}\mathrm{Sr}\) remains unsplit, while the lowest excited term \(5\mathrm{s}5\mathrm{p}\,^3\mathrm{P}\) splits into the
\(^3\mathrm{P}_0\), \(^3\mathrm{P}_1\), and \(^3\mathrm{P}_2\) fine‐structure levels.

Hyperfine structure results from the interaction of \(\mathbf{J}\) with the nuclear spin given by the total nuclear angular momentum \(\mathbf{I}\) resulting in the total atomic angular momentum
\begin{equation}
  \mathbf{F} = \mathbf{J} + \mathbf{I}\,,  
\end{equation}
where
\begin{equation}
  |J - I| \le F \le J + I\,.  
\end{equation}
However, \(^{88}\mathrm{Sr}\) has nuclear spin \(I=0\). As a result, its level structure exhibits only fine‐structure splittings, without the additional multiplet substructure seen in fermionic isotopes. Consequently, all hyperfine constants, the magnetic dipole \(A\), electric quadrupole \(B\), magnetic octupole \(C\), etc., vanish for this isotope.

\subsection{Interaction with Magnetic Fields}

The interaction between static external fields and $^{88}$Sr produces energy shifts and state couplings that are central in precision‐metrology, quantum‐simulation, and quantum computing protocols. In this section, we first examine how a static magnetic field lifts the degeneracy of the $m_J$ sublevels via the Zeeman effect. Then, we discuss state mixing, both intrinsic (spin–orbit) and externally induced, which enables otherwise forbidden transitions.

\subsubsection{Zeeman Effect}

In $^{88}$Sr, which has no nuclear spin and thus no hyperfine structure, atomic energy levels are characterized by their total electronic angular momentum $J$. Each state with angular momentum $J$ contains $2J + 1$ magnetic sublevels corresponding to the magnetic quantum number $m_J$, describing the projection of $\vec{J}$ along the quantization axis.

In the absence of an external magnetic field, these $m_J$ sublevels are energetically degenerate. When a magnetic field is applied, this degeneracy is lifted due to the Zeeman effect. The Hamiltonian describing the interaction of the atomic electrons with a static magnetic field $\vec{B}$, aligned along the $z$-axis which defines the quantization axis, is
\begin{equation}
H_B = \frac{\mu_{\rm B}}{\hbar} \left( g_S \vec{S} + g_L \vec{L} \right) \cdot \vec{B}
= \frac{\mu_{\rm B}}{\hbar} \left( g_S S_z + g_L L_z \right) B_z~,
\label{eq:hamiltonian_magnetic_field}
\end{equation}
where $\mu_{\rm B}$ is the Bohr magneton, $g_S$ and $g_L$ are the electron spin and orbital $g$-factors respectively, and $S_z$ and $L_z$ are the spin and orbital angular momentum components along the $z$-axis.

The value for $g_S$ has been measured precisely, and we list the CODATA recommended value in Tab.~\ref{tab:sr_physical_properties}. The orbital $g$-factor can be corrected for finite nuclear mass by~\cite{BetheSalpeter1957}
\begin{equation}
g_L = 1 - \frac{m_{\rm e}}{m_\text{nuc}},
\end{equation}
where $m_{\rm e}$ is the electron mass and $m_\text{nuc}$ is the nuclear mass, accurate to leading order in $m_{\rm e}/m_\text{nuc}$. 

When the energy shifts due to the magnetic field are small compared to the fine-structure splitting, $J$ remains a good quantum number and the Hamiltonian simplifies to
\begin{equation}
H_B = \frac{\mu_{\rm B}}{\hbar} g_J J_z B_z~.
\label{eq:zeeman_hamiltonian}
\end{equation}
Here, the Landé $g$-factor $g_J$ is given by
\begin{equation}
g_J = g_L \frac{J(J + 1) - S(S + 1) + L(L + 1)}{2J(J + 1)} 
+ g_S \frac{J(J + 1) + S(S + 1) - L(L + 1)}{2J(J + 1)}~.
\end{equation}
Using $g_S \approx 2$ and $g_L \approx 1$, $g_J$ can be approximated as
\begin{equation}
g_J \approx 1 + \frac{J(J + 1) + S(S + 1) - L(L + 1)}{2J(J + 1)}~.
\end{equation}
Note that the expression here does not include corrections due to the complicated multi-electron structure~\cite{BetheSalpeter1957}. The operator \(J_z\) in Eq.~\ref{eq:zeeman_hamiltonian} acts as
\begin{equation}
     J_z\,\lvert J,m_J\rangle \;=\; \hbar\,m_J\,\lvert J,m_J\rangle~.
\end{equation}
Hence, to first order, the energy shift for each Zeeman sublevel is
\begin{equation}
    \Delta E_{J,m_J} = \bra{J, m_J} H_B \ket{J, m_J} = \frac{\mu_{\rm B}}{\hbar} g_J (\hbar m_J) B_z = \mu_{\rm B} g_J m_J B_z~.
    \label{eq:energy_shift}
\end{equation}
The corresponding frequency shift is
\begin{equation}
    \Delta\omega \;=\; \frac{\Delta E}{\hbar}
\;=\;
\frac{\mu_{\rm B}}{\hbar}\,g_J\,m_J\,B_z~.
\end{equation}
This linear Zeeman effect leads to an equal spacing between adjacent magnetic sublevels, proportional to the magnetic field strength and the state's $g_J$ factor. These shifts play a central role in magneto-optical trapping, laser spectroscopy, and atomic clock operation using strontium.

\subsubsection{State Mixing}

In bosonic \textsuperscript{88}Sr, there is no hyperfine interaction to mix states. All state mixing arises from spin–orbit coupling in the $5\mathrm{s}5\mathrm{p}$ manifold.
The mixing can be described using the methods in Refs.~\cite{BreitWills1933, Lurio1962, Boyd2007}. The physical states $^{2S+1}L_J$ are written as admixtures of pure spin-orbit coupling states $^{2S+1}L_J^{(0)}$~\cite{Boyd2007}:
\begin{equation}
\begin{pmatrix} |\,^1\mathrm{P}_1\rangle \\[6pt] |\,^3\mathrm{P}_0\rangle \\[6pt] |\,^3\mathrm{P}_1\rangle \\[6pt] |\,^3\mathrm{P}_2\rangle \end{pmatrix} = \begin{pmatrix} \alpha & 0 & -\beta & 0 \\[4pt] 0 & 1 & 0 & 0 \\[4pt] \beta & 0 & \alpha & 0 \\[4pt] 0 & 0 & 0 & 1 \end{pmatrix} \, \begin{pmatrix} |\,^1\mathrm{P}_1^{(0)}\rangle \\[6pt] |\,^3\mathrm{P}_0^{(0)}\rangle \\[6pt] |\,^3\mathrm{P}_1^{(0)}\rangle \\[6pt] |\,^3\mathrm{P}_2^{(0)}\rangle \end{pmatrix}~,
\end{equation}
with $\alpha^2+\beta^2=1$.  For Sr one finds $\alpha \approx 0.9996$ and $\beta \approx -0.0286$~\cite{Boyd2007}. Hence, the $\lvert\,^3\mathrm{P}_1\rangle$ state carries a $\beta^2 \approx 0.08\%$ singlet admixture.
The ground state $\lvert\,^1\mathrm{S}_0\rangle$ is a pure singlet state, and electric‐dipole (E1) selection rules would forbid a direct $\;^1\mathrm{S}_0\to\,^3\mathrm{P}_J$ transition if the excited states were pure. However, since
\begin{equation}
   \bra{^1\mathrm{S}_0} \hat{\mathbf{d}} \ket{^3\mathrm{P}_1}
=\beta\,
 \bra{^1\mathrm{S}_0}\hat{\mathbf{d}} \ket{^1\mathrm{P}_1^{(0)}}
\neq0 ~,
\end{equation}
the \SI{689}{\nano\meter} intercombination transition becomes weakly allowed with a spontaneous decay rate $\propto\beta^2$, yielding the long lifetime $\tau_{3\mathrm{P}1}$. Here, $\hat{\mathbf{d}}$ is the dipole operator.

The $\;^1\mathrm{S}_0\to\,^3\mathrm{P}_0$ clock transition is strictly forbidden except via a two‐photon E1M1 process, giving an intrinsic lifetime of order $10^3\,$years~\cite{Santra2004}. More generally, one can relate the $^3\mathrm{P}_0$ lifetime to that of $^3\mathrm{P}_1$ by~\cite{BoydThesis2007}
\begin{equation}
\tau_{3\mathrm{P}0}
=
\Bigl(\frac{\omega_{3\mathrm{P}1}}{\omega_{3\mathrm{P}0}}\Bigr)^{\!3}
\frac{\beta^2}
{(\,\tilde\alpha_0\,\beta+\tilde\beta_0\,\alpha\,)^2}
\;\tau_{3\mathrm{P}1}~,
\label{eq:tau3P0}
\end{equation}
where $\tilde\alpha_0,\tilde\beta_0$ are the hyperfine‐mixing coefficients.  In bosonic $^{88}$Sr, $\tilde\alpha_0=\tilde\beta_0=0$, so the denominator vanishes and $\tau_{3\mathrm{P}0}\to\infty$ (i.e.\ $\sim10^3\,$yr), making the clock line unusable without external mixing.

Several proposals to realize such a mixing can be found in the literature, see e.g. discussion in Ref.~\cite{BoydThesis2007}. To make the $^1\mathrm{S}_0\to\,^3\mathrm{P}_0$ transition usable, a static magnetic field $B$ can be applied to mix the $\lvert\,^3\mathrm{P}_0\rangle$ state with $\lvert\,^3\mathrm{P}_1\rangle$ state via the magnetic‐dipole operator $\hat\mu$~\cite{taichenachev2006}
\begin{equation}
\lvert\,^3\mathrm{P}_0'\rangle
=
\lvert\,^3\mathrm{P}_0\rangle
+
\frac{\langle\,^3\mathrm{P}_1|\hat\mu\!\cdot\!B|\,^3\mathrm{P}_0\rangle}{\Delta_{10}}\,
\lvert\,^3\mathrm{P}_1\rangle,
\label{eq:Bmix}
\end{equation}
where $\Delta_{10}\equiv\omega_{3P1} - \omega_{3P0} \approx2\pi \times \SI{5.6}{\tera\hertz}$, see values in Tab.~\ref{tab:red_transition_data} and Tab.~\ref{tab:clock_transition_data}. 
Since $\lvert\,^3\mathrm{P}_1\rangle$ itself carries the small $\beta$‐admixture of $\lvert\,^1\mathrm{P}_1\rangle$, the doubly‐mixed clock state becomes
\begin{equation}
\lvert\,^3P_0'\rangle
= 
\lvert\,^3P_0\rangle
+
\frac{\langle\,^3P_1|\hat\mu\!\cdot\!B|\,^3P_0\rangle}{\Delta_{10}}
\bigl(\alpha\,\lvert\,^3P_1^{(0)}\rangle
     +\;\beta\,\lvert\,^1P_1^{(0)}\rangle\bigr)~.
\label{eq:Bmix_clock_full}
\end{equation}
The singlet component gives the otherwise forbidden $\;^1\mathrm{S}_0\to\,^3\mathrm{P}_0$ transition a nonzero E1 matrix element. The Hamiltonian in Eq.~\ref{eq:hamiltonian_magnetic_field} has no first-order energy shift on the clock state $\ket{^3\mathrm{P}_0}$, but it mixes it with the $\ket{^3\mathrm{P}_1, m_J = 0}$ state. This magnetic admixture induces a small but finite decay rate for the dressed clock state to the ground state~\cite{Madjarov2021}
\begin{equation}
\Gamma_{\mathrm{clock}}(\vec{B}) \,=\,
\Gamma_{\rm 3P1}
\cdot
\frac{\mu_C^2 \vec{B}^2}{2\Delta_{10}^2}~,
\end{equation}
where $\mu_C \equiv \sqrt{\frac{2}{3}}(g_L - g_S)\mu_{\rm B}$. Even for strong fields $\left|\vec{B}\right| \approx \SI{1000}{G}$, the induced linewidth remains narrow, e.g., $\Gamma_{\mathrm{clock}} \approx 2\pi \times \SI{0.3}{\milli\hertz}$ (i.e.\ $\tau \approx \SI{530}{\second}$)~\cite{Madjarov2021}. This magnetic mechanism is essential for enabling precision spectroscopy on the bosonic clock transition. Notably, the same transition in fermionic $^{87}$Sr is weakly allowed even at $B = 0$ due to hyperfine-induced state mixing, which can be understood as an effective internal magnetic field arising from the nonzero nuclear spin $I = 9/2$.

\section{Transition Properties}

This section examines the key atomic transitions. Measured transition frequencies are presented and converted to wavelengths and transition energies. Experimentally determined lifetimes of the excited states are reviewed. Equations for obtaining additional parameters from these transition data and lifetimes are introduced.

\subsection{Transition Frequencies and Wavelengths}

The following relations allow one to convert a transition frequency $\omega_0 = 2 \pi \times \nu$ to the corresponding transition energy, vacuum wavelength, air wavelength, and wavenumber. The photon energy $E$ is
\begin{equation}
  E \;=\; h\,\nu \;=\; \hbar\,\omega_0 ~.
\end{equation}
This transition energy can be converted into an energy in eV using
\begin{equation}
  E_{\mathrm{eV}} \;=\; \frac{h\,\nu}{e} \;=\; \frac{\hbar\,\omega_0}{e}\,.
\end{equation}
The vacuum wavelength \(\lambda\) and the angular wavenumber $k_\mathrm{L}$ are
\begin{align}
  \lambda \; &=\; \frac{c}{\nu}\, \;=\; \frac{2\pi\,c}{\omega_0}~,\\
  k_\mathrm{L} \; &=\; \frac{2\pi}{\lambda} = \frac{\omega_0}{c}\,.
\end{align}
The wavelength in air is given by
\begin{equation}
  \lambda_{\rm air} = \frac{\lambda}{n_{\rm air}}\,,
\end{equation}
where ${n_{\rm air}}$ is the refractive index of air. To compute the refractive index of air under typical laboratory conditions, the Edlén’s equation is a common approximation~\cite{Edlen1966,Peck1972} 
\begin{equation}
n_{\rm air} = 1 
+
\Biggl[
\bigl(8342.54 + \tfrac{2\,406\,147}{130 - \kappa^2} + \tfrac{15\,998}{38.9 - \kappa^2}\bigr)
\frac{P}{96\,095.43}
\frac{1 + 10^{-8}(0.601 - 0.009\,72\,T)\,P}{1 + 0.003\,661\,0\,T}
-
f\,(0.037\,345 - 0.000\,401\,\kappa^2)
\Biggr]\times10^{-8},
\label{Edlen1993}
\end{equation}
where \(\kappa = k_L/(2\pi)\) is the vacuum wavenumber in \(\mu\mathrm{m}^{-1}\), \(P\) is the total air pressure in Pa, \(T\) is the temperature in \(^\circ\mathrm{C}\), \(f\) is the partial pressure of water vapor in Pa (obtainable from relative humidity via the Buck equation~\cite{Buck1981, BuckResearch2012}). Under standard conditions ($P=\SI{101325}{\pascal}$, $T=\SI{25}{\celsius}$, \SI{50}{\percent} relative humidity), this yields \(n_{\rm air}\approx 1.00027\) in the visible range with a $1\sigma$ uncertainty of $1\times10^{-8}$ which has been used throughout this reference. In real‐world applications, additional uncertainties due to fluctuations in pressure, temperature, and humidity must be considered and are not included here, see e.g.\ Refs.~\cite{Steck2024SodiumDLine, Steck2024Rb85DLine, Steck2024Rb87DLine, Steck2024CesiumDLine}.

We now review the experimental measurements of the transition frequencies that we present in this work. For the \(^1\mathrm{S}_{0}\!\rightarrow{}^1\mathrm{P}_{1}\) blue-MOT transition, we combine a recent high-precision measurement~\cite{Lee2026} with the historical value reported in Ref.~\cite{Sansonetti2010}, which is based on a measurement from 1936~\cite{Sullivan1936}. Accurate frequency measurements of the \(^1\mathrm{S}_{0}\!\rightarrow{}^3\mathrm{P}_{1}\), \(^3\mathrm{P}_{1}\!\rightarrow{}^3\mathrm{S}_{1}\), and \(^3\mathrm{P}_{0}\!\rightarrow{}^3\mathrm{S}_{1}\) transitions were performed in 2005~\cite{Courtillot2005}. The absolute frequency of the \(^1\mathrm{S}_{0}\!\rightarrow{}^3\mathrm{P}_{0}\) clock transition in \(^{88}\mathrm{Sr}\) was experimentally measured and we obtain a weighted mean value of \SI{429228066418008(2)}{\hertz}~\cite{Baillard2007, Morzynski2015}. However, in Tab.~\ref{tab:clock_transition_data}, we adopt the value from the 2025 CIPM recommendation published by the BIPM~\cite{BIPM2025_88Sr429THz}, which is endorsed by the CCTF as a secondary representation of the SI second~\cite{CCTF2021PSFS2}. The absolute frequency of the magnetic quadrupole transition $^1\mathrm{S}_0 \rightarrow\,^3\mathrm{P}_2$ at $\lambda \approx \SI{671}{\nano\meter}$ was measured using cold atoms trapped in a lattice and an optical frequency comb~\cite{Trautmann2023}.

\noindent For the repump transition $^{3}\mathrm{P}_2 \to {}^{3}\mathrm{S}_1$, we infer the transition frequency in $^{88}\mathrm{Sr}$ from a measurement performed on the isotope $^{87}\mathrm{Sr}$. Specifically, we use the measured transition frequency $\nu_{87}$ between the hyperfine levels $^{3}\mathrm{P}_2(F=\tfrac{7}{2}) \to {}^{3}\mathrm{S}_1(F=\tfrac{7}{2})$ in $^{87}\mathrm{Sr}$, reported in 2019~\cite{Onishchenko2019}.
To convert this frequency to the corresponding transition frequency in $^{88}\mathrm{Sr}$, we first obtain the center–of–gravity frequency of the transition $\nu_{87}^{\rm CG}$ by removing the hyperfine shifts of the lower state \(\Delta_{3\mathrm{P}2(F=7/2)} = \SI{1597.138(8)}{\mega\hertz}\)~\cite{Heider1977} and the upper state \(\Delta_{3\mathrm{S}1(F=7/2)} = \SI{2981.0(6)}{\mega\hertz}\)~\cite{Courtillot2005}
\begin{equation}
  \nu_{87}^{\rm CG}
  = \nu_{87}
    - \bigl[\Delta_{3\mathrm{P}2(F=7/2)} - \Delta_{3\mathrm{S}1(F=7/2)}\bigr]\,.
\end{equation}
Then, we subtract the isotope shifts for the two levels, \(\delta_{\rm iso, 3\mathrm{S}1}\)~\cite{Courtillot2005} and \(\delta_{\rm iso, 3\mathrm{P}2}\)~\cite{Trautmann2023}, yielding the corresponding transition frequency in \({}^{88}\)Sr
\begin{equation}
  \nu_{88}
  = \nu_{87}^{\rm CG}
    + \bigl[\delta_{\rm iso, 3\mathrm{S}1} - \delta_{\rm iso, 3\mathrm{P}2}\bigr]\,.
\end{equation}

All strontium isotopes share the same electronic level structure, but differences in nuclear mass and charge distribution shift their optical transition frequencies. The mass‐dependent and volume‐dependent contributions combine to yield the total isotope shift for each line. The transition Tabs.~\ref{tab:blue_transition_data}-\ref{tab:m2_transition_data} compile published values of the isotope shift among the four stable Sr isotopes relative to $^{88}\mathrm{Sr}$. Since $^{87}\mathrm{Sr}$ has a hyperfine structure, we present the shifts for the center-of-gravity frequency of the transitions.

\subsection{Decay of Excited States}

An excited state \(\lvert e\rangle\) that decays to several lower levels \(\{\lvert g_i\rangle\}\) is described by Einstein \(A\)‐coefficients \(A_{e\to g_i}\).  The total spontaneous decay rate is
\begin{equation}
  \Gamma \;=\;\sum_i A_{e\to g_i}\,.
\end{equation}
The excited‐state lifetime \(\tau\) is directly related to the total spontaneous decay rate
\begin{equation}
  \tau = \frac{1}{\Gamma}\,.
\end{equation}
The corresponding full‐width at half‐maximum of the transition's line shape is
\begin{equation}
  \Delta\nu_{\rm FWHM} = \frac{\Gamma}{2\pi}\,.
\end{equation}
Each decay channel \(i\) carries a branching ratio
\begin{equation}
  \beta_i = \frac{A_{e\to g_i}}{\Gamma}\,,
\end{equation}
so its partial decay rate is
\begin{equation}
  \Gamma_{\rm partial} = \beta_i\,\Gamma = A_{e\to g_i}\,.
\end{equation}
Thus, combining a precise measurement of \(\tau\) with known \(\beta_i\) directly yields the individual \(A\)‐coefficients.

We now review experimental measurements and theoretical predictions of the lifetimes and branching ratios of relevant excited states in strontium.
We first discuss the lifetime of the \(^1\mathrm{P}_{1}\) state. The two recent measurements are mutually consistent~\cite{Heinz2020,Puljic2025} and also agree with an earlier, less precise measurement~\cite{Nagel2005}. Other earlier measurements show varying degrees of disagreement with these values, with one high-precision result being notably inconsistent~\cite{Lurio1964,KellyMathur1980,Yasuda2006}.
The \(^{1}\mathrm{P}_1\) state decays predominantly to the \(^{1}\mathrm{S}_0\) ground state. A recent direct measurement determined the \(^{1}\mathrm{P}_1\rightarrow{}^{1}\mathrm{D}_2\) decay rate to be \(\SI{5.3(5)e3}{\per\second}\)~\cite{Okamoto2026}. An earlier indirect determination relied on a theoretically calculated \(^{1}\mathrm{D}_2\rightarrow{}^{1}\mathrm{S}_0\) electric-quadrupole decay rate and is therefore not fully independent of theoretical input~\cite{Hunter1986}. We consequently adopt the recent direct experimental value. Using the total \(^{1}\mathrm{P}_1\) decay rate \(\Gamma_{^{1}\mathrm{P}_1}\), the corresponding branching fraction is~\cite{Okamoto2026}
\begin{equation}
    \beta_{^{1}\mathrm{P}_1\rightarrow{}^{1}\mathrm{D}_2}
    =
    \frac{A_{^{1}\mathrm{P}_1\rightarrow{}^{1}\mathrm{D}_2}}
         {\Gamma_{^{1}\mathrm{P}_1}}
    =
    2.8(3)\times10^{-5}
    =
    \frac{1}{3.6(3)\times10^4}~,
\end{equation}
which is the value shown in Fig.~\ref{fig:Sr_I_energy_levels_reference}.
For the \(^{1}\mathrm{P}_1\rightarrow{}^{1}\mathrm{S}_0\) branching fraction listed in Tab.~\ref{tab:blue_transition_data}, we additionally include the measured leakage channels to the \(5\mathrm{s}4\mathrm{d}\,{}^{3}\mathrm{D}_1\) and \(5\mathrm{s}4\mathrm{d}\,{}^{3}\mathrm{D}_2\) states, \(A_{^{1}\mathrm{P}_1\rightarrow{}^{3}\mathrm{D}_1} =\SI{66(6)}{\per\second}\) and \(A_{^{1}\mathrm{P}_1\rightarrow{}^{3}\mathrm{D}_2} =\SI{2.4(2)e2}{\per\second}\)~\cite{Okamoto2025}. The ground-state branching fraction is therefore calculated as
\begin{equation}
    \beta_{^{1}\mathrm{P}_1\rightarrow{}^{1}\mathrm{S}_0}
    =
    1-
    \frac{
        A_{^{1}\mathrm{P}_1\rightarrow{}^{1}\mathrm{D}_2}
        +A_{^{1}\mathrm{P}_1\rightarrow{}^{3}\mathrm{D}_1}
        +A_{^{1}\mathrm{P}_1\rightarrow{}^{3}\mathrm{D}_2}
    }{
        \Gamma_{^{1}\mathrm{P}_1}
    }
    =
    0.9999705(26)~.
\end{equation}
Other direct decay channels are neglected at the present level of precision.
Recently, the lifetime of the \(^{1}\mathrm{D}_2\) state was measured to be \(\SI{421.9(18)}{\micro\second}\)~\cite{Okamoto2026}. Combining this result with the earlier lifetime measurement~\cite{Husain1988} yields a weighted mean of \(\SI{421.6(18)}{\micro\second}\). Historically, the branching ratio for the decay from \(^1\mathrm{D}_2\) into the \(^3\mathrm{P}_1\) and \(^3\mathrm{P}_2\) states was estimated to be approximately \(2:1\)~\cite{Barker2015,Xu03,Bauschlicher1985}, while more recent calculations predicted branching fractions of \(80.4(56)\,\%\) and \(19.6(56)\,\%\), respectively~\cite{UDportal}.
A recent measurement found corresponding values of \(82(4)\,\%\) and \(17.7(4)\,\%\)~\cite{Okamoto2026}, which we use in Fig.~\ref{fig:Sr_I_energy_levels_reference}.

\noindent Multiple independent studies have reported measurements of the \(^3\mathrm{P}_1\) excited‐state lifetime~\cite{Nicholson2015, Havey1976, Drozdowski1997, Ma1968, Husain1988}. Except for the results in Ref.~\cite{Husain1988}, all results agree within their quoted uncertainties. A recent measurement achieves a relative precision well below one percent, representing the current benchmark for this lifetime~\cite{Nicholson2015}. For the lifetime of the $^3\mathrm{S}_1$ state, four independent and inconsistent measurements were performed~\cite{Brinkmann1969, Havey1977, Jonsson1984, Heinz2020}. The weighted mean of these lifetime values agrees with a recent theoretical calculation, which predicts a lifetime of \SI{13.92(7)}{\nano\second}~\cite{UDportal}. The \(^3\mathrm{S}_1\) state decays predominantly to the \(^3\mathrm{P}_{0,1,2}\) states, but there is a small leakage channel to the \(^1\mathrm{P}_1\) state with a branching ratio of \(1.31(22)\times10^{-4}\)~\cite{UDportal}.

\noindent The metastable \(^3\mathrm{P}_2\) state decays via a magnetic-quadrupole transition back to the ground state. The partial decay rate on this transition was recently measured~\cite{Kluesener2024}. However, faster processes, including magnetic-dipole decay into the nearby \(^3\mathrm{P}_1\) level and thermal (blackbody-induced) transfer to higher-lying triplet states, dominate under typical laboratory conditions. As a result, the measured total lifetime of the \(^3\mathrm{P}_2\) manifold is reduced to a few hundred seconds, reflecting the combined contributions of all available decay pathways~\cite{Derevianko2001, Yasuda2004}. Additionally, Raman scattering processes induced by, e.g., trapping light fields can further shorten the effective lifetime.

\subsection{Additional Transition Parameters}

The spontaneous emission rate \(\Gamma\) determines the relative intensity of a spectral line. For a \(J\to J'\) electric‐dipole transition, \(\Gamma\) is directly linked to the absorption oscillator strength \(f\) via~\cite{Corney1977}
\begin{equation}
  \Gamma \;=\;\frac{e^2\,\omega_0^2}{2\pi\varepsilon_0\,m_{\rm e}\,c^3}\,\frac{2J+1}{2J'+1}\;f~.
\end{equation}
A larger \(f\) (and hence larger \(\Gamma\)) implies a stronger, broader line. The spontaneous decay rate for a two‐level atom is related to the dipole matrix element $d$ via~\cite{Demtroder2008}
\begin{equation}
\Gamma  
\;=\; \frac{\omega_0^3\,|d|^2}{3\pi\varepsilon_0\,\hbar\,c^3}
\quad\Longrightarrow\quad
|d|^2 \;=\; \frac{3\pi\varepsilon_0\,\hbar\,c^3}{\omega_0^3}\,\Gamma~.
\label{eq:gamma_matrix_element}
\end{equation}
If we want the line‐specific moment $d_i$, we replace $\Gamma\to\Gamma_i=\beta_i\Gamma$, and find $|d_i|^2
   =\beta_i\,|d|^2$.
It is common to present reduced matrix elements $\langle J\|e\,r\|J'\rangle$, which are related to the dipole matrix element by the
Wigner–Eckart theorem
\begin{equation}
  \langle J\|e\,r\|J'\rangle
  \;=\;\sqrt{2J'+1}\;|d_i|
  \;=\;\sqrt{2J'+1}\;\sqrt{\frac{3\pi\varepsilon_0\,\hbar\,c^3}{\omega_0^3}\,\beta_i\Gamma}\,.
\label{eq:reduced_matrix_element}
\end{equation}
This expression is used to calculate the reduced matrix elements provided in Tabs.~\ref{tab:blue_transition_data}-\ref{tab:last_repump_data}, 
where we use theoretically calculated \(\beta_i\) from Ref.~\cite{UDportal}.

Absorbing or emitting a resonant photon of momentum \(\hbar k_L\) changes the velocity of an atom by the recoil velocity
\begin{equation}
  v_r = \frac{\hbar\,k_L}{m}~.
\end{equation}
The corresponding change in kinetic energy $\frac12 m v_r^2 \;=\;\frac{\hbar^2\,k_L^2}{2m}$ can be written as a recoil energy $E_{\rm rec}=\hbar\omega_r$, which defines the recoil frequency $\omega_r$
\begin{equation}
  E_{\rm rec} = \hbar\omega_r = \frac{\hbar^2\,k_L^2}{2m}
  \quad\Longrightarrow\quad
  \omega_r = \frac{\hbar\,k_L^2}{2m}\,.
  \label{eq:recoil_energy}
\end{equation}
An atom moving at velocity \(v \ll c\) sees the optical frequency \(\omega_\mathrm{L}\) shifted by a Doppler shift
\begin{equation}
  \Delta\omega_\mathrm{d} = \frac{v}{c}\,\omega_\mathrm{L}~.
\end{equation}
In particular, for \(v=v_r\) one finds \(\Delta\omega_d=2\,\omega_r\), showing that a single recoil kick corresponds to a twice as high Doppler shift change.

From these motions arise two key temperatures. The recoil temperature
\begin{equation}
  T_r = \frac{\hbar^2\,k_L^2}{m\,k_{\rm B}}
\end{equation}
is the temperature corresponding to an ensemble with a one-dimensional root-mean-square momentum of one photon recoil~\cite{Steck2024CesiumDLine}. Note that the single‐photon recoil energy $E_{\rm rec}$ (see Eq.~\ref{eq:recoil_energy}) can be expressed as a temperature via $k_{\rm B} T = E_{\rm rec}$, giving
\begin{equation}
  T = \frac{\hbar^2\,k_L^2}{2m\,k_{\rm B}}\,,
\end{equation}
which is exactly half of the rms‐defined recoil temperature $T_r$. The Doppler temperature
\begin{equation}
  T_\mathrm{D} = \frac{\hbar\,\Gamma}{2\,k_{\rm B}}
\end{equation}
is the lowest temperature that Doppler cooling of a two‐level atom can reach, set by the balance of Doppler cooling and recoil heating~\cite{Lett1988}. 
Since bosonic Sr has no hyperfine or Zeeman substructure in its \(^1\mathrm{S}_0\) ground state, sub‐Doppler cooling mechanisms are absent for both the \(^1\mathrm{S}_0\!\to\!{}^1\mathrm{P}_1\) and the \(^1\mathrm{S}_0\!\to\!{}^3\mathrm{P}_1\) transitions. Consequently, the minimum achievable temperatures in magneto-optical traps using these transitions are limited by the Doppler cooling limit (see discussion below and Tabs.~\ref{tab:blue_transition_data} and \ref{tab:red_transition_data}).

\subsection{Magneto-Optical Traps}

The magneto-optical trap (MOT) enables laser cooling and confinement of neutral atoms for a wide range of experiments in atomic, molecular, and optical physics. When laser cooling strontium, it is common to start by utilizing the broad \(^1\mathrm{S}_0\!\to\!{}^1\mathrm{P}_1\) “blue” transition at $\lambda \approx \SI{461}{\nano\meter}$.
With a natural linewidth on the order of tens of megahertz, this transition provides a large capture velocity, making it ideal for rapidly decelerating thermal atoms and loading them into a MOT~\cite{Xu03}.
This is crucial, as strontium atoms move at several hundred meters per second after sublimation in an oven that typically operates at temperatures above \SI{400}{\celsius}~\cite{Schioppo2012}.
In practice, however, the \(^1\mathrm{S}_0\!\to\!{}^1\mathrm{P}_1\) transition is not a cycling transition and atoms can decay via the \(^1\mathrm{D}_2\) state to the long-lived \(^3\mathrm{P}_2\) manifold.
To transfer the atoms out of this dark state it is common to repump through \({}^3\mathrm{S}_1\), using the $^3\mathrm{P}_2 \to ^3\mathrm{S}_1$ transition at $\lambda \approx \SI{707}{\nano\meter}$ and the $^3\mathrm{P}_0 \to ^3\mathrm{S}_1$ transition at $\lambda \approx \SI{679}{\nano\meter}$~\cite{Xu03}. Alternatively, atoms trapped in the metastable \(5\mathrm{s}5\mathrm{p}\,^3\mathrm{P}_2\) state can be repumped via the \(5\mathrm{s}\,n \mathrm{d}\,^3\mathrm{D}_{1,2,3}\, , n\in\{4,5,6\}\) manifolds~\cite{Mickelson2009, Stellmer2014}. Atoms can also be repumped from the \(^1\mathrm{D}_2\) state, where it was shown that repumping via the \(5\mathrm{s}8\mathrm{p}\,^1\mathrm{P}_1\) transition at $\lambda \approx \SI{448}{\nano\meter}$ can enhance the number of trapped atoms by a factor of \(12.0(6)\), six times larger than the enhancement obtained using the \(5\mathrm{s}4\mathrm{d}\,^1\mathrm{D}_2\!\to\!5\mathrm{s}6\mathrm{p}\,^1\mathrm{P}_1\) transition at $\lambda \approx \SI{717}{\nano\meter}$~\cite{Samland2019, Okamoto2025}.

This returns atoms to the cooling cycle through the \(^3\mathrm{P}_1\!\to\!{}^1\mathrm{S}_0\) decay channel. Using repumping typically increases the number of atoms in the MOT by a factor of $10-20$, depending on the loading rate~\cite{Nagel2008, Xu03}. If the repumpers are applied continuously throughout the blue MOT, it maximizes the loading rate, but the number of atoms in the blue MOT will be limited by light-assisted collisions. Without repumping, the atoms that decay to the \(m_J=+1,2\) sublevels of \(^3\mathrm{P}_2\) will be magnetically trapped, and the repumpers are only turned on for a few milliseconds toward the end of the blue MOT~\cite{Nagel2003}. This increases the overall number of trapped atoms, but it decreases the loading rate.

Once the atoms are cooled to the millikelvin regime in the blue MOT, they can be transferred into a second‐stage “red” MOT operating on the narrow \(^1\mathrm{S}_0\!\to\!{}^3\mathrm{P}_1\) intercombination transition at $\lambda \approx \SI{689}{\nano\meter}$.
This transition features a natural linewidth of only a few kHz with an associated Doppler limit of just a few hundred nanokelvin, enabling cooling well below the recoil limit of the blue transition~\cite{Katori1999}.
To improve the transfer of atoms between these two MOT stages, it is common to artificially broaden the linewidth of the red MOT laser~\cite{Katori1999} and rapidly ramp down the gradient of the magnetic coils~\cite{GonzalezEscudero2016, Norcia2017, Kristensen2023}. The red MOT typically operates with a gradient of $3-\SI{5}{G/cm}$, whereas the blue MOT operates with a gradient of $30-\SI{50}{G/cm}$~\cite{Katori1999,Xu03}. The broadening of the red MOT laser can be removed once the atoms have been captured from the blue MOT and cooled to the microkelvin regime.

In the red MOT, the scattering force from the radiation pressure becomes comparable to gravity when the red MOT beams are not frequency broadened. Since $^{88}\mathrm{Sr}$ has no hyperfine structure, the atoms in the red MOT will occupy only the surface of an ellipsoid defined by the Zeeman-shifted resonance condition~\cite{Stellmer2013_PRA}
\begin{equation}
    -\hbar\,\Delta \;=\; g_J\,\mu_{\rm B}\,B(\mathbf{r})~,
\end{equation}
where \(\Delta\) is the detuning of the MOT laser from the zero‐field resonance,  \(g_J\) is the Landé \(g\)-factor, \(\mu_{\rm B}\) is the Bohr magneton, \(B(\mathbf{r})\) is the magnitude of the quadrupole magnetic field at position \(\mathbf{r}\).
Inside this ellipsoidal shell the atoms experience no restoring force, and will accumulate in a thin “cap” at the bottom of the ellipsoid where the upward radiation‐pressure force from the MOT beams exactly compensates gravity~\cite{Loftus2004}. 

In practice, this means that the lowest achievable temperature in the red MOT is set by the requirement to hold atoms against gravity, typically yielding final temperatures on the order of a few microkelvin~\cite{Katori1999}.

\section[Electric Fields]{Interaction with Electric Fields}

The interaction between an atom and an external electric field underpins both static energy shifts and coherent optical manipulations in modern atomic‐physics experiments. We first derive the DC Stark effect, separating the induced energy shift into scalar, vector, and tensor contributions based on the atomic polarizability in a static field. These shifts give rise to state‐dependent optical dipole potentials and define special operating wavelengths, the so-called “magic,” “anti-magic,” and “tune-out” wavelengths. Finally, we address the dynamic response of the atom to near‐resonant light, deriving the Rabi frequency that governs coherent population oscillations under the rotating‐wave approximation.

\subsection{Polarizability}

In its most general form, the interaction Hamiltonian between an atom and a classical electric field is
\begin{equation}
  H_E = -\,\hat{\mathbf{d}}\cdot\mathbf{E}~,
\end{equation}
where $\hat{\mathbf{d}}$ is the electric dipole operator and $\mathbf{E}$ is the applied field. We choose the quantization axis along the $z$–direction, so that for a linearly polarized field is
\begin{equation}
  \mathbf{E} = E_z\,\hat{\mathbf{e}}_z~, 
  \quad
  H_E = -\,d_z\,E_z\,.
\end{equation}
Far from resonance, the amplitude of the dipole moment is related to the electric field amplitude via~\cite{HertelSchulz2015}
\begin{equation}
  d(\mathbf{r}) = \alpha(\omega)\,E(\mathbf{r})
\end{equation}
where $E(\mathbf{r})$ is the local field amplitude. Here
\begin{equation}
  \alpha(\omega) = \Re\{\alpha(\omega)\} \;+\; i\,\Im\{\alpha(\omega)\}
\end{equation}
is the (complex) dynamic polarizability at angular frequency $\omega$.  The real part $\Re\{\alpha(\omega)\}$ gives the energy shift, while the imaginary part $\Im\{\alpha(\omega)\}$ accounts for absorption and scattering.
Hence, the resulting (time-averaged) energy shift can be written as~\cite{Grimm2000}
\begin{equation}
  H_E \;\approx\; -\frac{1}{2}\,\Re\{\alpha(\omega)\}\,E_z^2 \,.
\end{equation}
In the following, we define the dynamic polarizability as $\alpha_{\rm tot}\,=\,\Re\{\alpha(\omega)\}$.
The dynamic polarizability of a state \( |i\rangle \) with angular momentum $J_i$ can be expressed as a sum of contributions from the scalar, vector, and tensor parts~\cite{LeKien2013, Tsyganok2019, Trautmann2023}
\begin{equation}
\label{eq:alpha_tot}
\begin{aligned}
\alpha_{\rm tot}^i(\omega) &= 
  \underbrace{\alpha_s^i(\omega)}_{\text{scalar}}
  + \underbrace{\alpha_v^i(\omega)\,\sin(2\gamma)\,\frac{m_{J_i}}{2J_i}}_{\text{vector}}
  + \underbrace{\alpha_t^i(\omega)\,\frac{3m_{J_i}^2 - J_i(J_i+1)}{2J_i\,(2J_i-1)}\,(3\cos^2\beta -1)}_{\text{tensor}}~,
\end{aligned}
\end{equation}
where $\gamma$ is the ellipticity angle of the light polarization and $\cos(\beta)$ is the projection of \textbf{E} onto the quantization axis. In the generalized form of the polarizability, it is necessary to include the contribution of multiple transitions as well as the frequency of the electric field, $\omega$. The scalar polarizability of an atomic state \( |i\rangle \) is~\cite{Trautmann2023}
\begin{equation}
\alpha_s^i(\omega) = \frac{2}{\hbar} \cdot \frac{1}{3(2J_i + 1)} \sum_k \frac{|\langle k \| \mathbf{d} \| i \rangle|^2 \omega_{ki}}{\omega_{ki}^2 - \omega^2}~,
\end{equation}
where \( \langle k \| \mathbf{d} \| i \rangle \) is the reduced electric dipole matrix element and \( \hbar\omega_{ki} = \hbar\omega_k - \hbar\omega_i \) is the energy difference between the excited state \( |k\rangle \) and the initial state \( |i\rangle \). The sum runs over all dipole-allowed transitions to states \( |k\rangle \).
The vector polarizability is given by~\cite{Trautmann2023}
\begin{equation}
\alpha_v^i(\omega)
= -\frac{2}{\hbar}\,
\sqrt{\frac{6J_i}{(J_i+1)(2J_i+1)}}
\sum_k(-1)^{J_i+J_k}
\begin{Bmatrix}
1 & 1 & 1\\
J_i & J_k & J_i
\end{Bmatrix}
\frac{\bigl|\langle k\|\mathbf d\|i\rangle\bigr|^2\, \omega}
     {\omega_{ki}^2-\omega^2}\,,
\end{equation}
where the curly brackets are the Wigner 6-j symbols.
Finally, the tensor polarizability is~\cite{Trautmann2023}
\begin{equation}
\alpha_t^{i}(\omega) = \frac{2}{\hbar}\,\sqrt{ \frac{10 J_i (2J_i - 1)}{3(J_i + 1)(2J_i + 1)(2J_i + 3)} } 
\sum_k (-1)^{J_i + J_k} 
\begin{Bmatrix}
1 & 2 & 1 \\
J_i & J_k & J_i
\end{Bmatrix}
\frac{|\langle k \| \mathbf{d} \| i \rangle|^2 \omega_{ki}}{(\omega_{ki}^2 - \omega^2)}~.
\end{equation}
The static polarizability can be found by setting $\omega=0$.

\subsection{Types of state-dependent potentials}

The optical dipole potential \(U\) experienced by an atom in a light field can be written, in the far‐off‐resonant, low‐saturation limit, as~\cite{Grimm2000}
\begin{equation}
      U(\mathbf{r})
  = -\frac{\alpha_{\rm tot}^i(\omega)}{\epsilon_0 c}\,I(\mathbf{r})\,,
\end{equation}
where \(I\) is the time‐averaged intensity.
A positive (negative) polarizability produces an attractive (repulsive) potential.
At certain “special” wavelengths the polarizabilities of two states satisfy particular relations, giving rise to state‐(in)dependent potentials:
\begin{itemize}
    \item A \emph{magic wavelength} \(\lambda_{\rm m}\) is defined by
\[
  \alpha_{\rm tot}^g(\omega_{\rm m}) = \alpha_{\rm tot}^e(\omega_{\rm m})\,,
\]
so that ground \((g)\) and excited \((e)\) states experience identical potentials.  This cancels differential Stark shifts and preserves the unperturbed transition frequency.
    \item An \emph{anti‐magic wavelength} \(\lambda_{\rm a}\) satisfies
\[
  \alpha_{\rm tot}^g(\omega_{\rm a}) = -\,\alpha_{\rm tot}^e(\omega_{\rm a})\,,
\]
so that the ground- and excited-state dipole potentials have equal magnitude but opposite sign, $U_g(\mathbf{r}) = -\,U_e(\mathbf{r})$.
    \item A \emph{tune‐out wavelength} \(\lambda_{\rm t}\) occurs when the polarizability of one state vanishes:
\[
  \alpha_{\rm tot}^g(\omega_{\rm t}^g)=0
  \quad\text{or}\quad
  \alpha_{\rm tot}^e(\omega_{\rm t}^e)=0~.
\]
At \(\lambda_{\rm t}\), an atom in one internal state experiences no optical potential while in the other it remains confined to a lattice site, enabling state-selective mobility and precise tunneling control.
\end{itemize}

Theoretical values for the magic, anti-magic, and tune-out wavelengths can be found in databases such as in Refs.~\cite{UDportal, Barakhshan2022}. Table~\ref{tab:special_wavelengths} presents experimentally measured magic and tune-out wavelengths, supplemented with theoretical values. Note that some values were not measured with the isotope $^{88}\mathrm{Sr}$ and the precise wavelengths could vary due to, e.g., isotope shifts. The tune-out wavelength for the \(^{1}\mathrm{S}_{0}\) ground state has been measured with high precision~\cite{Heinz2020}. Experimental studies of the \(^{1}\mathrm{S}_{0}\!\to\!{}^{3}\mathrm{P}_{1}(m_{J})\) intercombination line have revealed magic wavelengths that produce both attractive, red-detuned optical traps~\cite{Ido2003,Norcia2018,Kestler2022} and repulsive, blue-detuned traps~\cite{Kestler2023}.
Likewise, measurements of the \(^{1}\mathrm{S}_{0}\!\to\!{}^{3}\mathrm{P}_{0}\) clock transition have identified a blue-detuned magic wavelength near \(\SI{389.9}{\nano\meter}\)~\cite{Takamoto2009} and red-detuned magic wavelengths near \(\SI{476.8}{\nano\meter}\)~\cite{Ma2025}, \(\SI{497.4}{\nano\meter}\)~\cite{Kestler2025}, and \(\SI{813.4}{\nano\meter}\). All magic wavelengths are isotope dependent because the dynamic polarizabilities are determined by isotope-shifted transition frequencies. Consequently, measurements performed in one isotope cannot, in general, be transferred directly to another isotope. We therefore use only measurements that explicitly use \(^{88}\mathrm{Sr}\) to obtain the recommended values in this reference. Measurements performed in other isotopes are included for completeness and marked with an asterisk in Tab.~\ref{tab:special_wavelengths}. For the commonly used magic wavelength near \(\SI{813.4}{\nano\meter}\), we combine the direct \(^{88}\mathrm{Sr}\) determinations of Refs.~\cite{Takano2017,Origlia2018}. We exclude both the estimate of Ref.~\cite{Akatsuka2010}, which was obtained from an \(^{87}\mathrm{Sr}\) measurement using calculated isotope and hyperfine corrections, and measurements performed exclusively in
\(^{87}\mathrm{Sr}\)~\cite{Takamoto2005,Wang2016,Ushijima2018, Kim2023,Jia2026}.
The magic wavelength near $\sim\SI{813.4}{\nano\meter}$, widely used in optical lattice clocks, can be magic for both the clock and the \(^{1}\mathrm{S}_{0}\!\to\!{}^{3}\mathrm{P}_{1}\) transition at a suitable polarization angle, thereby enhancing the performance of direct sideband cooling of atoms trapped in tweezers or optical lattices~\cite{Norcia2019}. Recently, a “triple-magic” condition, where the \(^{1}\mathrm{S}_{0}\), \(^{3}\mathrm{P}_{0}\), and \(^{3}\mathrm{P}_{2}(m_{J}=0)\) states experience identical trapping potentials, was demonstrated by tuning the polarization of the red-detuned light to a magic angle~\(\beta\)~\cite{Ammenwerth2024}. At $\beta=\SI{90}{\degree}$, theoretical calculations predict a magic wavelength for the $^3\mathrm{P}_0 \to {}^3\mathrm{P}_2(m_J=0)$ transition of $\lambda_{\rm m} = \SI{1010}{\nano\meter}$ with an uncertainty of $\pm\SI{171}{\nano\meter}$~\cite{UDportal}. Recent measurements for these states observed a magic condition at $\lambda_{\rm m} =\SI{914}{\nano\meter}$ for a polarization angle close to the theoretically predicted angle of $\beta=\SI{79}{\degree}$, but found no magic condition at $\lambda=\SI{1064}{\nano\meter}$~\cite{Pucher2024}.

\subsection{Calculation of Rabi Frequencies}
\label{sec:calc_rabi_general}

Consider an atom with a ground‐state manifold $\ket{g} =  \ket{J_g,\,m_{J_g}}$ and an excited‐state manifold $\ket{e} =  \ket{J_e,\,m_{J_e}}$. A near‐resonant laser with an angular frequency $\omega_L$ drives the $\ket{g}  \leftrightarrow \ket{e}$ transition via the electric‐dipole Hamiltonian:
\begin{equation}
    H_{\mathrm{int}}(t)
\;=\;
-\,\hat{\mathbf{d}}\;\cdot\;\mathbf{E}(t) 
\;\approx\; 
-\,\bra{e} \hat{\mathbf{d}}\cdot\hat{\boldsymbol{\epsilon}}_{\,q} \ket{g} 
\;E_{0}\,\cos(\omega_{L} t)\;\bigl(\ket{e} \bra{g} + \ket{g}\bra{e}\bigr)~,
\end{equation}
where $\bigl\langle e \bigl|\hat{\mathbf{d}}\cdot\hat{\boldsymbol{\epsilon}}_{\,q}\bigr| g \bigr\rangle$ is the matrix element of the dipole operator between the two levels, including the polarization component.
Here, \(q\) labels the spherical polarization component. \(q=0\) corresponds to \(\pi\)-polarization, i.e., electric field along the quantization axis, while \(q=+1\) and \(q=-1\) correspond to right-circular \(\sigma^{+}\) and left-circular \(\sigma^{-}\) polarization, respectively.
Under the rotating‐wave approximation, the Rabi frequency $\Omega_{q}^{\,g\to e}$ is
\begin{equation}
\Omega_{q}^{\,g\to e} 
\;=\; 
\frac{1}{\hbar}\,\bra{e}\hat{\mathbf{d}}\cdot\hat{\boldsymbol{\epsilon}}_{\,q} \ket{g} \;E_{0}.
\label{eq:Omega_basic_general}
\end{equation}
Applying the Wigner–Eckart theorem~\cite{HertelSchulz2015}, one can separate the total matrix element into a product of a reduced dipole matrix element $\langle e\,\|\,\hat{\mathbf{d}}\,\|\,g\rangle$ (which depends on $J_g$ and $J_e$ but not on $m_{J_g}$ or $m_{J_e}$) and a Clebsch–Gordan factor that accounts for the coupling between the specific Zeeman sublevels $(m_{J_g},\,m_{J_e})$ under polarization $q$. The Clebsch–Gordan coefficient can be calculated using the Wigner $3$‐$j$ symbol
\begin{equation}
\langle\,J_g\,m_{J_g};\,1\,q \mid J_e\,m_{J_e}\rangle
= (-1)^{-J_g+1-m_{J_e}}\sqrt{2J_e+1}
\begin{pmatrix}
J_g & 1 & J_e \\
m_{J_g} & q & -m_{J_e}
\end{pmatrix}~.
\label{eq:clebsch_gordan}
\end{equation}
This separation greatly simplifies practical calculations, since all the dependence on magnetic quantum numbers and polarization is contained in the known tabulated Clebsch–Gordan factor. The reduced matrix element of Sr can be found in the literature or in a database~\cite{UDportal}.
Explicitly, the separation gives~\cite{Auzinsh2010}
\begin{equation}
\bra{e} \hat{\mathbf{d}}\cdot\hat{\boldsymbol{\epsilon}}_{\,q} \ket{g} 
\;=\; 
(-1)^{-J_g+1-m_{J_e}}\;\begin{pmatrix}
  J_g & 1 & J_e \\[3pt]
  m_{J_g} & q & -\,m_{J_e}
\end{pmatrix}
\;\langle e\,\|\,\hat{\mathbf{d}}\,\|\,g\rangle
\;\equiv\; 
w_{q}^{\,g\to e}\;\langle e\,\|\,\hat{\mathbf{d}}\,\|\,g\rangle~.
\label{eq:wigner_factor_def}
\end{equation}
Note that here the factor \(\sqrt{2J_e+1}\) that appears in the relation between the Clebsch–Gordan coefficients and the Wigner $3$‐$j$ symbols is absorbed in the definition of the reduced matrix element, see Eq.~\ref{eq:reduced_matrix_element}. Substituting into Eq.~\ref{eq:Omega_basic_general} yields the (real, positive) Rabi frequency
\begin{equation}
\Omega_{q}^{\,g\to e} 
\;=\; 
\bigl|w_{q}^{\,g\to e}\bigr| \frac{\bigl|\langle e\,\|\,\hat{\mathbf{d}}\,\|\,g\rangle\bigr|}{\hbar}\;E_{0}~.
\label{eq:Omega_separated}
\end{equation}

We consider an atom driven by a Gaussian laser beam, whose waist is much larger than the spatial extent of the atomic wavefunction. In this case, the atom experiences the peak intensity of the Gaussian beam, given by
\begin{equation}
    I_{0}
    =\;\frac{2P}{\pi w_{0}^{2}}
    \;=\;
    \frac{8P}{\pi D^{2}}\,,
\end{equation}
where \(P\) is the total optical power and \(w_{0}\) is the \(1/e^{2}\) beam radius so that the full \(1/e^{2}\) diameter is \(D=2w_{0}\).
Using Eq.~\ref{eq:intensity_efield}, the peak electric‐field amplitude is
\begin{equation}
I_{0}
=\;\frac12\,c\,\varepsilon_{0}\,E_{0}^{2}
\quad\Longrightarrow\quad
E_{0}
=\;\sqrt{\frac{2I_{0}}{\varepsilon_{0}\,c}}
=\;\sqrt{\frac{16\,P}{\varepsilon_{0}\,c\,\pi\,D^{2}}}\,.
\label{eq:E0_gaussian_peak}
\end{equation}

Using the separated form of the Rabi frequency (see Eq.~\ref{eq:Omega_separated}) together with the peak electric‐field amplitude (see Eq.~\ref{eq:E0_gaussian_peak}), we find  
\begin{equation}
\label{eq:Omega_final_gaussian_clean}
\Omega_{q}^{\,g\to e}
=\;
\bigl|w_{q}^{\,g\to e}\bigr|\;\frac{\bigl|\langle e\,\|\,\hat{\mathbf{d}}\,\|\,g\rangle\bigr|}{\hbar}
\;\sqrt{\frac{16\,P}{\varepsilon_{0}\,c\,\pi\,D^{2}}}~.
\end{equation}
In Tables~\ref{tab:blue_transition_data}-\ref{tab:m2_transition_data}, we list numerical values of \(\Omega_{q}^{\,g\to e}\) obtained from Eq.~\ref{eq:Omega_final_gaussian_clean} under the following common experimental parameters: a laser power of \(P=1\:\mathrm{mW}\) with a Gaussian beam diameter of \(D=1\:\mathrm{mm}\), and $\pi$‐polarization (\(q=0\)). Specifically, we calculate the Rabi frequency for \(m_J=0\to m'_J=0\) transitions. For the \(^3\mathrm{P}_1\to{}^3\mathrm{S}_1\) transition, the \(m_J=0\to m'_J=0\) transition is dipole‐forbidden. Instead, we evaluate the Rabi frequency for the allowed \(\sigma^+\) transition \(m_{J_g}=0\to m_{J_e}=+1\) (\(q=+1\)).

We now discuss the calculation of the Rabi frequency on the $^1\mathrm{S}_0\,\rightarrow\,^3\mathrm{P}_0$ clock transition. In Eq.~\ref{eq:Bmix}, we discussed the mixing of the $^3\mathrm{P}_0$ state with the $^3\mathrm{P}_1$ state. The resulting induced transition rate, i.e., the on-resonance Rabi frequency, is~\cite{taichenachev2006}
\begin{equation}
\Omega_{\rm clock}
=\frac{1}{\hbar^2\,\Delta_{10}}
\;\langle\,^1\mathrm{S}_0|\hat{\mathbf{d}}\!\cdot\!\vec{E}|\,^3\mathrm{P}_1\rangle\;
\langle\,^3\mathrm{P}_1|\boldsymbol{\hat\mu}\!\cdot\!\vec{B}|\,^3\mathrm{P}_0\rangle
\;\propto\;
\left|\vec{B}\right|\,\sqrt{I_{\rm clock}}~.
\label{eq:rabi_clock}
\end{equation}
where $I_{\rm clock}$ is the clock‐laser intensity. The clock Rabi frequency can be derived to be~\cite{Madjarov2021}
\begin{equation}
\Omega_{\rm clock}(B) = \frac{\mu_C B}{\hbar \Delta_{10}} \sqrt{ \frac{3 \Gamma_{\rm 3P1} \lambda^3 I_{\rm clock}}{4\pi^2 \hbar c} }\, (\hat{\epsilon} \cdot \vec{B})~,
\label{eq:rabi_madjarov}
\end{equation}
where $\lambda$ is the clock transition wavelength, $\hat{\epsilon} \cdot \vec{B}$ is the projection of the clock beam polarization onto the magnetic field axis, and $\mu_C$ and $\Delta_{10}$ as defined previously. The constant terms can be combined in a single value $\alpha = 2 \pi \times 201.9(2)\,\frac{\mathrm{Hz}}{\rm T\sqrt{\mathrm{(mW/cm^2)}}}$~\cite{taichenachev2006}. Then, the Rabi frequency is given by
\begin{equation}
\Omega_{\rm clock}\,=\,\alpha\,\sqrt{I_{\rm clock}}\,|\vec{B}|\,\cos\theta~,
\end{equation}
where $\theta$ is the angle between the clock-laser polarization and the magnetic field.

The beam that excites the clock transition also couples to other transitions and hence induces an AC Stark shift on both the ground state $|\,^1\mathrm{S}_0\rangle$ and the clock state $|\,^3\mathrm{P}_0\rangle$. Since both states have total angular momentum $J=0$, their dynamic polarizabilities are purely scalar, and hence the induced shift is independent of the probe polarization. The differential light shift of the clock transition is therefore simply proportional to the intensity of the clock beam
\begin{equation}
\Delta\omega_I(I_{\rm clock})
\;=\;
\kappa\,I_{\rm clock}~.
\end{equation}
where $\kappa \;=\;-2\pi\times18\;\mathrm{mHz}\,/(\mathrm{mW}/\mathrm{cm}^2)\,$~\cite{taichenachev2006, Madjarov2021}.
Because the Rabi frequency scales as $\Omega_{\rm clock}\propto\sqrt{I_{\rm clock}}$ (cf. Eq.~\ref{eq:rabi_clock}), the relative light shift per Rabi oscillation grows as
\begin{equation}
\frac{\Delta\omega_I}{\Omega}
\;\propto\;
\frac{I}{\sqrt{I}}
\;=\;
\sqrt{I}\,.
\end{equation}
In practice, this means that pushing to higher clock Rabi frequencies by increasing the clock beam intensities inevitably amplifies the differential light shift faster than it improves the Rabi rate. Thus, higher clock beam intensities demand correspondingly better intensity stabilization to keep $\Delta\omega_I$ under control.
Since the clock states have $J=0$, there is no first‐order Zeeman shift of the $^1S_0\to{}^3P_0$ transition. To second order in the magnetic‐dipole perturbation one finds~\cite{taichenachev2006, Madjarov2021}
\begin{equation}
\Delta\omega_B(\vec{B})
=-\frac{\mu_C^2}{\,\Delta_{10}}\,\vec{B}^2
\;\equiv\;\eta\,\vec{B}^2,
\end{equation}
where \(\eta\approx -2\pi\times\SI{0.23}{\hertz/G\squared}\)~\cite{Madjarov2021}. This quadratic Zeeman shift is one of the dominant systematics in bosonic Sr clocks~\cite{Origlia2018}. Its magnitude scales as $\vec{B}^2$, so operating at lower bias fields (at the expense of reduced Rabi rate $\Omega\propto \left|\vec{B}\right|$) can mitigate Zeeman‐induced inaccuracy.

At sufficiently high magnetic fields, mixing of the $^3\mathrm{P}_2$ state also gives the $^1\mathrm{S}_0\rightarrow{}^3\mathrm{P}_2$ transition an electric dipole character similar to the clock transition. However, in contrast to the clock transition, the $^1\mathrm{S}_0 \rightarrow\,^3\mathrm{P}_2$ transition can be driven even without magnetic field, because it is magnetic-quadrupole (M2) allowed~\cite{Trautmann2023}. The corresponding Rabi frequency depends on the laser’s magnetic‐field amplitude (i.e., beam intensity), its propagation direction relative to the quantization axis, and the light’s polarization. Note that in the case of M2 transitions, magnetic quadrupole selection rules apply instead of the standard dipole selection rules. A full derivation of the M2 Rabi frequency is beyond the scope of this review. Instead, we adopt in Tab.~\ref{tab:m2_transition_data} the experimentally measured value from Ref.~\cite{Kluesener2024} and refer the reader to Ref.~\cite{Trautmann2023} for further details.

\section[Fluorescence]{Resonance Fluorescence}
\label{sec:resonance_fluorescence}

In this section, we analyze the interaction of a two‐level atom with a monochromatic classical electromagnetic field. Under the electric‐dipole and rotating‐wave approximations, the atom behaves as an ideal two‐level system whose dynamics are governed by the optical Bloch equations. We derive these equations, solve for the steady‐state excited‐state population, and then obtain expressions for the total photon scattering rate, the saturation intensity, and the on‐resonance scattering cross section.

\subsection{Optical Bloch Equations}

Consider an atom with ground state $\lvert g \rangle$ and excited state $\lvert e \rangle$, separated by an energy $\hbar \omega_0$. We drive the $\lvert g \rangle \leftrightarrow \lvert e \rangle$ transition with a classical, linearly polarized, monochromatic field of the form
\begin{equation}
\mathbf{E}(t) \;=\; E_0 \, \hat{\boldsymbol{\epsilon}} \, \cos(\omega_L t) \,,
\end{equation}
where $E_0$ is the real field amplitude, $\hat{\boldsymbol{\epsilon}}$ is the unit polarization vector, and $\omega_L$ is the laser frequency. The atom–field interaction Hamiltonian in the electric‐dipole approximation is
\begin{equation}
H_{\rm int} \;=\; -\,\hat{\mathbf{d}} \,\cdot\, \mathbf{E}(t) \;=\; -\,\bigl(d\,\hat{\sigma}_+ + d^*\,\hat{\sigma}_-\bigr)\,E_0 \cos(\omega_L t)\,,
\end{equation}
where $\hat{\mathbf{d}}$ is the atomic dipole moment, $d = \langle e \lvert \hat{\mathbf{d}}\cdot \hat{\boldsymbol{\epsilon}} \lvert g \rangle$ is the (complex) dipole matrix element, and $\hat{\sigma}_- = \lvert g \rangle \langle e \rvert$ (with $\hat{\sigma}_+ = \hat{\sigma}_-^\dagger$) is the lowering operator for the two‐level system. We define the Rabi frequency as
\begin{equation}
\Omega \;=\; \frac{d \, E_0}{\hbar}\,.
\label{eq:rabi}
\end{equation}
We move into a frame rotating at the laser frequency $\omega_L$ and apply the rotating‐wave approximation (RWA). Then, the density operator $\rho$ of the two‐level atom is described by the optical Bloch equations~\cite{Steck2024CesiumDLine}
\begin{subequations}
\label{eq:optical_bloch_raw}
\begin{align}
\dot{\rho}_{gg} \; &= \; \frac{i \Omega}{2} \bigl(\rho_{eg} - \rho_{ge}\bigr) \;+\; \Gamma\,\rho_{ee} \,, 
\\
\dot{\rho}_{ee} \; &= \; -\,\frac{i \Omega}{2} \bigl(\rho_{eg} - \rho_{ge}\bigr) \;-\; \Gamma\,\rho_{ee} \,,
\\
\dot{\rho}_{ge} \; &= \; -\bigl(\gamma + i \Delta\bigr)\,\rho_{ge} \;-\; \frac{i \Omega}{2}\,\bigl(\rho_{ee} - \rho_{gg}\bigr) \,,
\end{align}
\end{subequations}
where $\rho_{ij} = \langle i \lvert \rho \rvert j \rangle$ are the matrix elements of the density operator in the $\{\lvert g \rangle, \lvert e \rangle\}$ basis, $\gamma = \tfrac{\Gamma}{2} + \gamma_c$ is the total coherence‐decay (``transverse'') rate, with $\gamma_c$ a pure‐dephasing rate (e.g., due to collisions), and $\Delta = \omega_L - \omega_0$ is the detuning of the driving field from atomic resonance.

In writing Eqs.~\ref{eq:optical_bloch_raw}, we have neglected any additional couplings to auxiliary levels or motional effects. The term $\Gamma\,\rho_{ee}$ in the population equations accounts for radiative decay from $\lvert e \rangle$ to $\lvert g \rangle$, while the coherence $\rho_{ge}$ and its conjugate $\rho_{eg} = \rho_{ge}^*$ evolve under both dephasing and the coherent drive $\Omega$.

\subsection{Steady‐State Excited‐State Population}

Here, we are primarily interested in the long‐time (steady‐state) solution, where $\dot{\rho}_{ij} = 0$ for all $i,j$. In the steady state and assuming the purely radiative case ($\gamma = \Gamma/2$, i.e., $\gamma_c = 0$), one finds
\begin{equation}
\rho_{ee}^{(\infty)}
\;=\;
\frac{\displaystyle \bigl(\tfrac{\Omega}{\Gamma}\bigr)^2}{\displaystyle 1 + 4\bigl(\tfrac{\Delta}{\Gamma}\bigr)^2 + 2\bigl(\tfrac{\Omega}{\Gamma}\bigr)^2}
~.
\label{eq:rho_ee_ss}
\end{equation}
We can rewrite this equation
\begin{equation}
\rho_{ee}^{(\infty)} \;=\; \frac{s/2}{1 + s + 4(\Delta/\Gamma)^2}\,, 
\end{equation}
where $s$ is the on‐resonance saturation parameter, given by
\begin{equation}
s \;=\; \frac{2\Omega^2}{\Gamma^2}~.
\end{equation}
Equation~\ref{eq:rho_ee_ss} shows that, on resonance ($\Delta=0$), the excited‐state population saturates to $1/2$ when $\Omega \gg \Gamma$.

\subsection{Total Photon Scattering Rate}

The total photon scattering (or fluorescence) rate is the rate at which population decays from the excited state, given by
\begin{equation}
    R_{\rm sc} \;=\; \Gamma \,\rho_{ee}~.
\end{equation}
In the steady state, we find
\begin{equation}
R_{\rm sc} 
\;=\; 
\Gamma \,\rho_{ee}^{(\infty)} 
\;=\; 
\frac{\Gamma}{2} \; \frac{\,s\,}{\,1 + s + 4(\Delta/\Gamma)^2\,} \,,
\label{eq:R_sc_raw}
\end{equation}
where we have set $s = 2\Omega^2/\Gamma^2$ as above. Rewriting Eq.~\ref{eq:R_sc_raw} in terms of the incident intensity $I$ leads naturally to the definition of the saturation intensity $I_{\rm sat}$.

\subsection{Saturation Intensity}

The average intensity of a plane electromagnetic wave in free space is 
\begin{equation}
    I \;=\; \frac{1}{2}\,c \,\varepsilon_0 \, E_0^2\,,
    \label{eq:intensity_efield}
\end{equation}
where $c$ is the speed of light and $\varepsilon_0$ is the vacuum permittivity. In terms of the Rabi frequency (see Eq.~\ref{eq:rabi}), one finds
\begin{equation}
I 
= 
\frac{1}{2} \, c \,\varepsilon_0 \,\Bigl(\hbar \,\frac{\Omega}{\lvert d \rvert}\Bigr)^2 
= 
\frac{c\,\varepsilon_0\,\hbar^2}{2 \lvert d \rvert^2}\;\Omega^2 \,.
\end{equation}
We define the (on‐resonance) saturation intensity $I_{\rm sat}$ using the condition that $s=1$ when $I = I_{\rm sat}$. Since $s = 2\Omega^2 / \Gamma^2$, setting $s=1$ yields
\begin{equation}
I_{\rm sat} 
\;=\; 
\frac{c\,\varepsilon_0\,\hbar^2}{2 \lvert d \rvert^2}\;\frac{\Gamma^2}{2}
\;=\; 
\frac{c\,\varepsilon_0\,\hbar^2\,\Gamma^2}{4\,\lvert d \rvert^2}\,.
\label{eq:I_sat}
\end{equation}
Using Eq.~\ref{eq:gamma_matrix_element}, we find the saturation intensity for a cycling transition
\begin{equation}
I_{\rm sat}
= \frac{\hbar\,\omega_0^3\,\Gamma}{12\pi\,c^2}
= \frac{\pi\,h\,c\,\Gamma}{3\,\lambda^3}\,.
\end{equation}
Equivalently, $I_{\rm sat}$ and $\Omega$ are often defined as
\begin{equation}
\frac{I}{I_{\rm sat}}
=\frac{2\,\Omega^2}{\Gamma^2}
=s
\quad\Longrightarrow\quad
\Omega
=\frac{\Gamma}{\sqrt2}\sqrt{\frac{I}{I_{\rm sat}}}~.
\label{eq:saturation_parameter}
\end{equation}
Using this definition of $I_{\rm sat}$, we can write the scattering rate (see Eq.~\ref{eq:R_sc_raw}) as
\begin{equation}
R_{\text{sc}} = \left( \frac{\Gamma}{2} \right) \frac{\left( \frac{I}{I_{\text{sat}}} \right)}{1 + 4 \left( \frac{\Delta}{\Gamma} \right)^2 + \left( \frac{I}{I_{\text{sat}}} \right)}~.
\label{eq:gain_equation}
\end{equation}
When the laser is exactly on resonance ($\Delta = 0$), the scattering rate reduces to
\begin{equation}
R_{\rm sc}(\Delta=0) 
= 
\frac{\Gamma}{2} \;\frac{\,I / I_{\rm sat}\,}{\,1 + I / I_{\rm sat}\,}
\,.
\label{eq:R_sc_on_res}
\end{equation}
Thus, on resonance, the scattering rate on a fully saturated transition ($I \gg I_{\rm sat}$) approaches $\Gamma/2$, consistent with the fact that at saturation $\rho_{ee}^{(\infty)} \to 1/2$.

\subsection{On‐Resonance Scattering Cross Section}

It is often convenient to express the scattering properties in terms of an effective cross section $\sigma(\Delta, I)$, defined as the power radiated by the atom divided by the incident energy flux so that the power scattered by the atom is $\sigma(\Delta, I)\,I$. Using Eq.~\ref{eq:gain_equation}, we can calculate the scattering cross section as
\begin{equation}
R_{\rm sc} \;=\; \sigma(\Delta, I)\, I \quad\Longrightarrow\quad
\sigma(\Delta, I)
\;=\;
\frac{R_{\rm sc}}{\,I\,}
\;=\;
\frac{\Gamma}{2\,I}\;\frac{I / I_{\rm sat}}{\,1 + I/I_{\rm sat} + 4(\Delta/\Gamma)^2\,}
\,.
\end{equation}
We define the low-intensity on‐resonance cross section as
\begin{equation}
\sigma_0 \;=\; \frac{\hbar \omega_0\,\Gamma}{2\,I_{\rm sat}} 
~.
\label{eq:sigma_0}
\end{equation}
For a cycling transition, this reduces to~\cite{Foot2004}
\begin{equation}
\sigma_0 \;=\; \frac{3 \lambda^2}{2\pi}
~.
\label{eq:sigma_0_two_level}
\end{equation}
Then, we can write the scattering cross section as
\begin{equation}
\sigma(\Delta, I) 
= 
 \;\frac{\sigma_0}{\,1 + 4 (\Delta / \Gamma)^2 + I / I_{\rm sat}\,}\,.
\label{eq:sigma_general}
\end{equation}
On resonance ($\Delta = 0$), this reduces to
\begin{equation}
\sigma(0, I) 
= 
\frac{\sigma_0}{\,1 + I / I_{\rm sat}\,}\,,
\label{eq:sigma_on_res}
\end{equation}
showing that at $I = I_{\rm sat}$ the scattering cross section is half of its low‐intensity value $\sigma_0$.

\clearpage

\section{Data Tables}

\begin{table}[h!]
\renewcommand{\arraystretch}{1.15} 
\centering
\caption{Fundamental Physical Constants (2022 CODATA recommended values)~\cite{Mohr2025}.}
\label{tab:fundamental_constants}
\begin{tabular}{|l|c|c|}
\hline
\textbf{Constant}            & \textbf{Symbol} & \textbf{Value}                                                   \\ \hline
Speed of Light               & $c$             & \SI{2.99792458e8}{\metre\per\second} (exact)                    \\ \hline
Permeability of Vacuum       & $\mu_{0}$       & \SI{1.25663706127(20)e-6}{\newton\per\ampere\squared}                \\ \hline
Permittivity of Vacuum       & $\varepsilon_{0}$ & $(\mu_{0}c^{2})^{-1}\approx$ \SI{8.8541878188(14)e-12}{\farad\per\metre}                 \\ \hline
Planck’s Constant            & $h$             & \SI{6.62607015e-34}{\joule\second} (exact)                     \\ \hline
Reduced Planck Constant      & $\hbar$         & $\sim$ \SI{1.054571817e-34}{\joule\second}                    \\ \hline
Elementary Charge            & $e$             & \SI{1.602176634e-19}{\coulomb} (exact)                         \\ \hline
Bohr Magneton                & $\mu_{\rm B}$       & \SI{9.2740100657(29)e-24}{\joule\per\tesla}                    \\ \hline
Atomic Mass Unit             & $u$             & \SI{1.66053906892(52)e-27}{\kilogram}                          \\ \hline
Electron Mass                & $m_{\rm e}$         & \SI{9.1093837139(28)e-31}{\kilogram}                           \\ \hline
Bohr Radius                  & $a_{0}$         & \SI{5.29177210544(82)e-11}{\metre}                            \\ \hline
Boltzmann Constant           & $k_{\rm B}$         & \SI{1.380649e-23}{\joule\per\kelvin} (exact)                   \\ \hline
\end{tabular}
\end{table}

\begin{table}[h!]
\centering
\caption{Physical properties of strontium-88.}
\renewcommand{\arraystretch}{1.15} 
\label{tab:sr_physical_properties}
\begin{tabular}{|l|c|c|c|}
\hline
\textbf{Property} & \textbf{Symbol} & \textbf{Value} & \textbf{Reference} \\
\hline
Atomic Number & $Z$ & 38 & \\
\hline
Total Nucleons & $Z + N$ & 88 & \\
\hline
Nuclear Spin & $I$ & 0 & \\
\hline
Particle Statistics &  & bosonic & \\
\hline
Relative Natural Abundance & $\eta\mathopen{}\left({}^{88}\mathrm{Sr} \right)$ & 82.58(1)\% & \cite{crc97} \\
\hline
Nuclear Lifetime & $\tau_n$ & stable & \\
\hline
Atomic Mass & $m$ & \makecell[c]{
\SI{87.905612254(6)}{u}\\
\SI{1.4597070353(5)e-25}{\kilogram}} & \cite{Wang2021} \\
\hline
RMS Nuclear Charge Radius & $R$ & \SI{4.2240(18)}{\femto\meter} & \cite{Angeli2013} \\
\hline
Electron Spin g-Factor  & $g_S$ & \SI{2.00231930436092(36)}{} & \cite{Mohr2025} \\
\hline
Electron Orbital g-Factor & $g_L$ & $\sim$ \SI{0.99999376}{} & \\
\hline
Density at \SI{20}{\celsius} & $\rho_m$ & \SI{2582(4)}{\kilogram /\cubic\metre} & \cite{Arblaster2018} \\
\hline
Molar Volume at \SI{20}{\celsius} & $V_m$ & \SI{33.94(5)}{\centi \cubic\metre /\mol} & \cite{Arblaster2018} \\ 
\hline
Melting Point & $T_{\rm M}$ & \SI{1050}{\kelvin} (\SI{777}{\celsius}) & \cite{crc97} \\
\hline
Boiling Point & $T_{\rm B}$ & \SI{1655}{\kelvin} (\SI{1382}{\celsius}) & \cite{crc97} \\
\hline
Specific Heat Capacity & $c_p$ & \SI{0.306}{\joule\per\gram\per\kelvin} & \cite{crc97} \\
\hline
Molar Heat Capacity & $C_p$ & \SI{26.79}{\joule\per\mol\per\kelvin} & \cite{crc97} \\
\hline
Vapor Pressure at \SI{25}{\celsius} & $P$ & \makecell[c]{
  \SI{3.39(17)e-18}{\pascal}\\
  \SI{2.54(13)e-20}{}~Torr
}  & \cite{alcock1984vapor} \\
\hline
Ionization Limit & $E_{\rm I}$ & \makecell[c]{
  $45\,932.09(15)\,\mathrm{cm}^{-1}$\\
  \SI{5.694853(19)}{\electronvolt}
} & \cite{Rubbmark1978} \\
\hline
\end{tabular}
\end{table}

\begin{table}[h!]
\renewcommand{\arraystretch}{1.15} 
  \centering
  \caption{Scattering lengths between the strontium isotopes, given in units of~$a_0$.}
  \begin{tabular}{|l|c|c|c|c|}
    \hline
      & $^{84}\mathrm{Sr}$ & $^{86}\mathrm{Sr}$ & $^{87}\mathrm{Sr}$ & $^{88}\mathrm{Sr}$ \\
    \hline
    $^{84}\mathrm{Sr}$ & 122.7(3)~\cite{MartinezDeEscobar2008, Stein2008, Stein2010}  & 31.9(3)~\cite{MartinezDeEscobar2008, Stein2008, Stein2010}   & $-56.1(10)$~\cite{MartinezDeEscobar2008, Stein2008, Stein2010} & 1768(124)~\cite{MartinezDeEscobar2008, Stein2010} \\ \hline
    $^{86}\mathrm{Sr}$ &  & 810.6(9)~\cite{Mickelson2005, Ferrari2006, MartinezDeEscobar2008, Stein2008, Stein2010, Aman2018}  & 162.5(5)~\cite{MartinezDeEscobar2008, Stein2008, Stein2010}   &  97.40(10)~\cite{Ferrari2006, MartinezDeEscobar2008,Stein2008, Stein2010}  \\ \hline
    $^{87}\mathrm{Sr}$ &  &  &  96.20(10)~\cite{MartinezDeEscobar2008, Stein2008, Stein2010}   &  54.98(19)~\cite{MartinezDeEscobar2008, Stein2008, Stein2010}  \\ \hline
    $^{88}\mathrm{Sr}$ &  &  &  &  $-0.8(18)$ ($R_{\mathrm{B}}: 3.1$)~\cite{Mickelson2005, Yasuda2006, Ferrari2006, MartinezDeEscobar2008, Stein2008, Stein2010}\\ \hline
  \end{tabular}
  \label{tab:sr_scattering_lengths}
\end{table}


\begin{table}[h!]
\centering
\renewcommand{\arraystretch}{1.15}
\caption{Landé $g_J$ factors and Zeeman coefficients $\mu_{\rm B} g_J$ for selected states of $^{88}$Sr.}
\begin{tabular}{|l|c|c|}
\hline
\textbf{State} & \boldmath$g_J$ & \boldmath$\mu_{\rm B} g_J$ (MHz/G) \\
\hline
$^1\mathrm{S}_0$ & 0  & 0 \\\hline
$^1\mathrm{P}_1$ & 1  & 1.39962 \\\hline
$^3\mathrm{P}_0$ & 0  & 0 \\\hline
$^3\mathrm{P}_1$& 1.50116  & 2.10106 \\\hline
$^3\mathrm{P}_2$& 1.50116  & 2.10106 \\\hline
$^3\mathrm{S}_1$& 2.00232  & 2.80249 \\
\hline
\end{tabular}
\label{tab:zeeman_coeffs}
\end{table}


\begin{table}[h!]
\renewcommand{\arraystretch}{1.2}
\setlength{\tabcolsep}{2pt}
\centering
\caption{Experimentally measured magic and tune‐out wavelengths, supplemented with theoretical polarizability values. Asterisks denote values measured using strontium isotopes other than $^{88}\mathrm{Sr}$.}
\label{tab:special_wavelengths}
\begin{tabular}{|l|l|c|c|c|c|}
\hline
\textbf{Type} & \textbf{Transition} & $\boldsymbol{\beta}$ (°) & $\boldsymbol{\lambda}$ (nm) & $\boldsymbol{\omega_0/2\pi}$ (THz) & $\boldsymbol{\alpha}_{\rm tot}$ (a.u.) \\
\hline
\multirow{13}{*}{magic}
  & \multirow{3}{*}{$^1\mathrm{S}_0\!-\!^3\mathrm{P}_0$}                          
    & $\forall \angle$     & $389.889(9)^{*}$~\cite{Takamoto2009}      & $768.92(2)$           & $-459$~\cite{Witkowski2022}       \\ \cline{3-6}
  &
    & $\forall \angle$     & $476.82362(8)$~\cite{Ma2025}      & $628.7282(1)$~\cite{Ma2025}           & $2818$~\cite{UDportal}       \\ \cline{3-6}
  & 
    & $\forall \angle$     & $497.4363(3)^{*}$~\cite{Kestler2025}      & $602.67507(36)$           & $1337$~\cite{UDportal}       \\ \cline{2-6}
    & $^1\mathrm{S}_0\!-\!^3\mathrm{P}_0\!-\!^3\mathrm{P}_1(m_J=0)$
    & $\sim 49$~\cite{Norcia2019}
    & \multirow{2}{*}{$813.427\,156(26)$}
    & \multirow{2}{*}{%
        \makecell[c]{
        $368.554\,769(12)$\\
        ($R_{\mathrm B}: 1.84$)~\cite{Takano2017,Origlia2018}}}
    & \multirow{2}{*}{$286$~\cite{UDportal}}
    \\ \cline{2-3}

  & $^1\mathrm{S}_0\!-\!^3\mathrm{P}_0\!-\!^3\mathrm{P}_2(m_J=0)$
    & $78.49(3)$~\cite{Ammenwerth2024}
    & & &
    \\ \cline{2-6}
  & \multirow{3}{*}{$^1\mathrm{S}_0\!-\!^3\mathrm{P}_1(m_J=0)$}
    & $0$           & $473.371(6)$~\cite{Kestler2022}       & $633.314(8)$ & $3573(16)$ \cite{Kestler2022}    \\ \cline{3-6}
  &                                                                
    & $24(1)$~\cite{Norcia2018} & $\sim 515.13$~\cite{Norcia2018} & $\sim 581.97$ & $949$~\cite{UDportal} \\ \cline{3-6}
  &                                                                
    & $90$          & $914(1)$ \cite{Ido2003}               & $328.0(4)$            & $261$~\cite{UDportal}            \\ \cline{2-6}
  & \multirow{2}{*}{$^1\mathrm{S}_0\!-\!^3\mathrm{P}_1(m_J=1)$}
    & $0$           & $435.827(25)$ \cite{Kestler2023}      & $687.87(4)$           & $-1558(57)$~\cite{Kestler2023}   \\ \cline{3-6}
  &                                                                
    & $0$           & $473.117(15)$ \cite{Kestler2022}      & $633.65(2)$           & $3637(17)$~\cite{Kestler2022}    \\ \cline{2-6}
  & \multirow{2}{*}{$^3\mathrm{P}_0\!-\!^3\mathrm{P}_2(m_J=0)$}                    
    & $\sim 9.5$ \cite{Unnikrishnan2024}         & $\sim 539.91$ \cite{Unnikrishnan2024}           & $\sim 555.26$             & $431$~\cite{UDportal}            \\ \cline{3-6}
  &
    & $90$
    & $997(1)$~\cite{Ammenwerth2026}
    & $300.7(3)$
    & $190$~\cite{UDportal} \\ \cline{2-6}
  & ${}^3\mathrm{P}_2(m_J=2) \!- \!5\mathrm{s}4\mathrm{d}\,{}^3\mathrm{D}_3(m_J=3)$ & $0$ & $597.14(3)$~\cite{Holman2026} & $502.047(25)$ & $386$~\cite{UDportal} \\ \cline{2-6}
\hline
tune‐out
  & $^1\mathrm{S}_0$                                            
    & $\forall \angle$     & $689.222\,22(2)$                         & $434.972\,130(10)$ \cite{Heinz2020}  & $0$                         \\
\hline
\end{tabular}
\end{table}

\begin{table}[h!]
\renewcommand{\arraystretch}{1.15}
\centering
\caption{$^1\mathrm{S}_0 \rightarrow\,^1\mathrm{P}_1$ Transition (Blue MOT).}
\begin{tabular}{|l|c|c|c|}
\hline
\textbf{Parameter} & \textbf{Symbol} & \textbf{Value} & \textbf{Ref.} \\ \hline
Frequency & $\omega_{0}$ &  \makecell{$2\pi\times$ \SI{650.503815(6)}{\tera\hertz} \\ $R_{\mathrm{B}}$: 1.2}  & \cite{NIST_ASD, Sullivan1936, Lee2026} \\ \hline
Transition Energy & $\hbar\omega_0$ & \SI{2.69026761(3)}{\electronvolt} & \\ \hline
Wavelength (vacuum) & $\lambda$ & \SI{460.861952(4)}{\nano\meter} & \\ \hline
Wavelength (air) & $\lambda_{\rm air}$ & \SI{460.737479(6)}{\nano\meter} & \\ \hline
Wavenumber (vacuum) & $k_L/2\pi$ & \SI{21698.4716(2)}{\centi\meter^{-1}} & \\ \hline
Lifetime excited state & $\tau$ & \makecell{$5.253(12)\,\mathrm{ns}$ \\ $R_{\mathrm{B}}$: 3.4} & \cite{Lurio1964, KellyMathur1980, Nagel2005, Yasuda2006, Heinz2020, Puljic2025} \\ \hline
Decay rate & $\Gamma$ & \SI{190.4(4)e6}{\second^{-1}} & \\ \hline
Natural Line Width & $\Delta\nu_{\rm FWHM}$ & $2\pi\times$ \SI{30.30(7)}{\mega\hertz} & \\ \hline
Branching Ratio & $\beta$ & \SI{0.9999705(26)}{} & \cite{Okamoto2025, Okamoto2026} \\ \hline
Oscillator strength & $f$ & 1.818(4) &  \\ \hline
Recoil velocity & $v_r$ & $\sim$ \SI{9.8496}{\milli\meter/\second} &  \\ \hline
Recoil energy & $\omega_r$ & $\sim 2\pi\times$ \SI{10.686}{\kilo\hertz} &  \\ \hline
Recoil temperature & $T_r$ & $\sim$ \SI{1026}{\nano\kelvin} &  \\ \hline
Doppler shift ($v=v_r$) & $\Delta\omega_d$ & $\sim 2\pi\times$\SI{21.372}{\kilo\hertz} &  \\ \hline
Doppler Temperature  & $T_\mathrm{D}$ & \SI{727(2)}{\micro\kelvin} &  \\ \hline
Reduced E1 Matrix Element & $\braket{J \| er \| J'}$ & $5.252(6) ea_0$ & \\ \hline
Saturation Intensity & $I_{\rm sat}$ & \SI{40.45(9)}{\milli\watt/\centi\meter\squared} &  \\ \hline
Resonant Cross Section & $\sigma_0$ & $\sim$ \SI{1.014e-13}{\meter\squared} &  \\ \hline
\makecell[l]{
  Rabi Frequency \\
\footnotesize $P=\SI{1}{\milli\watt}$, $D=\SI{1}{\milli\meter}$
} & $\Omega_0^{\ket{m_J=0} \text{\scalebox{0.7}[1]{$\to$}} \ket{m'_J=0}}$ & $2\pi\times$ \SI{53.75(6)}{\mega\hertz} &  \\ \hline
Isotope Shifts & 
\makecell{$\nu({}^{84}\mathrm{Sr})-\nu({}^{88}\mathrm{Sr})$\\
          $\nu({}^{86}\mathrm{Sr})-\nu({}^{88}\mathrm{Sr})$\\
          $\nu({}^{87}\mathrm{Sr})-\nu({}^{88}\mathrm{Sr})$} & 
\makecell{\SI{-275(3)}{\mega\hertz} ($R_{\mathrm{B}}$: 12.5)\\
          \SI{-126(2)}{\mega\hertz} ($R_{\mathrm{B}}$: 10.7)\\
          \SI{-47(2)}{\mega\hertz} ($R_{\mathrm{B}}$: 7.2)}
& \cite{Bushaw2000, Zhang2025} \\ \hline
\end{tabular}
\label{tab:blue_transition_data}
\end{table}

\begin{table}[h!]
\renewcommand{\arraystretch}{1.15} 
\centering
\caption{$^1\mathrm{S}_0 \rightarrow\,^3\mathrm{P}_1$ Intercombination Transition (Red MOT).}
\begin{tabular}{|l|c|c|c|}
\hline
\textbf{Parameter} & \textbf{Symbol} & \textbf{Value} & \textbf{Ref.} \\ \hline
Frequency & $\omega_0$ & $2\pi\times$ \SI{434.829 121 309(9)}{\tera\hertz} & \cite{Ferrari2003, Courtillot2005} \\ \hline
Transition Energy & $\hbar\omega_0$ & \SI{1.79830875068(4)}{\electronvolt} &  \\ \hline
Wavelength (vacuum) & $\lambda$ & \SI{689.448896839(14)}{\nano\meter} & \\ \hline
Wavelength (air) & $\lambda_{\rm air}$ & \SI{689.265521(7)}{\nano\meter} &  \\ \hline
Wavenumber (vacuum) & $k_L/2\pi$ & \SI{14504.3382415(3)}{\centi\meter^{-1}} & \\ \hline
Lifetime excited state & $\tau$ & \makecell{\SI{21.28(4)}{\micro\second} \\ $R_{\mathrm{B}}$: 1.4} & \cite{Ma1968, Havey1976, Husain1988, Drozdowski1997, Nicholson2015} \\ \hline
Decay rate & $\Gamma$ & \SI{47.00(9)e3}{\second^{-1}} &  \\ \hline
Natural Line Width & $\Delta\nu_{\rm FWHM}$ & $2\pi\times$ \SI{7.481(15)}{\kilo\hertz} & \\ \hline
Oscillator strength & $f$ & \SI{0.001005(2)}{} &  \\ \hline
Recoil velocity & $v_r$ & $\sim$ \SI{6.584}{\milli\meter/\second} &  \\ \hline
Recoil energy & $\omega_r$ & $\sim2\pi\times$ \SI{4.775}{\kilo\hertz} &  \\ \hline
Recoil temperature & $T_r$ & $\sim$ \SI{458}{\nano\kelvin} &  \\ \hline
Doppler shift ($v=v_r$) & $\Delta\omega_d$ & $\sim2\pi\times$ \SI{9.5496}{\kilo\hertz} &  \\ \hline
Doppler Temperature  & $T_\mathrm{D}$ & \SI{179.5(4)}{\nano\kelvin} &  \\ \hline
Reduced E1 Matrix Element & $\braket{J \| er \| J'}$ & \SI{0.15102(15)}{}\,$ea_0$ &  \\ \hline
Saturation Intensity & $I_{\rm sat}$ & \SI{2.983(6)}{\micro\watt/\centi\meter\squared} &  \\ \hline
Resonant Cross Section & $\sigma_0$ & $\sim$ \SI{2.2696e-13}{\meter\squared} &  \\ \hline
\makecell[l]{
  Rabi Frequency \\
\footnotesize $P=\SI{1}{\milli\watt}$, $D=\SI{1}{\milli\meter}$
} & $\Omega_0^{\ket{m_J=0} \text{\scalebox{0.7}[1]{$\to$}} \ket{m'_J=0}}$ & $2\pi\times$ \SI{1.5454(15)}{\mega\hertz} &  \\ \hline
Isotope shifts & 
\makecell{$\nu({}^{84}\mathrm{Sr})-\nu({}^{88}\mathrm{Sr})$\\
          $\nu({}^{86}\mathrm{Sr})-\nu({}^{88}\mathrm{Sr})$\\
          $\nu({}^{87}\mathrm{Sr})-\nu({}^{88}\mathrm{Sr})$} & 
\makecell{\SI{-351.496(5)}{\mega\hertz}\\
          \SI{-163.8174(2)}{\mega\hertz}\\
          \SI{-62.186(12)}{\mega\hertz}}& 
\makecell{\cite{Miyake2019}\\
          \cite{Miyake2019, Ferrari2003}\\
          \cite{Courtillot2005, Miyake2019}} \\ \hline
\end{tabular}

\label{tab:red_transition_data}
\end{table}

\begin{table}[h!]
\renewcommand{\arraystretch}{1.15} 
\centering
\caption{$^3\mathrm{P}_0 \rightarrow\,^3\mathrm{S}_1$ Repump Transition (\SI{679}{\nano\meter}).}
\begin{tabular}{|l|c|c|c|}
\hline
\textbf{Parameter} & \textbf{Symbol} & \textbf{Value} & \textbf{Ref.} \\ \hline
Frequency & $\omega_0$ & $2\pi\times$ \SI{441.3327513(7)}{\tera\hertz}  & \cite{Courtillot2005}  \\ \hline
Transition Energy & $\hbar\omega_0$ & \SI{1.825205603(3)}{\electronvolt} &  \\ \hline
Wavelength (vacuum) & $\lambda$ & \SI{679.288943(1)}{\nano\meter} &  \\ \hline
Wavelength (air) & $\lambda_{\rm air}$ & \SI{679.108203(7)}{\nano\meter} & \\ \hline
Wavenumber (vacuum) & $k_L/2\pi$ & \SI{14721.27599(2)}{\centi\meter^{-1}} &  \\ \hline
Lifetime excited state & $\tau$ & \makecell{ \SI{13.89(21)}{\nano\second} \\ $R_{\mathrm{B}}$: 1.9} & \cite{Brinkmann1969, Havey1977, Jonsson1984, Heinz2020} \\ \hline
Decay rate & $\Gamma$ & \SI{72.0(11)e6}{\second^{-1}} &  \\ \hline
Natural Line Width & $\Delta\nu_{\rm FWHM}$ & $2\pi\times$ \SI{11.46(17)}{\mega\hertz} & \\ \hline
Branching Ratio & $\beta$ & \SI{0.1161(10)}{} & \cite{UDportal} \\ \hline
Partial decay rate & $\Gamma_{\rm partial}$ & \SI{8.36(14)e6}{\second^{-1}} &  \\ \hline
Recoil velocity & $v_r$ & $\sim$ \SI{6.682}{\milli\meter/\second} &  \\ \hline
Recoil energy & $\omega_r$ & $\sim2\pi\times$ \SI{4.919}{\kilo\hertz} &  \\ \hline
Recoil temperature & $T_r$ & $\sim$ \SI{472}{\nano\kelvin} &  \\ \hline
Doppler shift ($v=v_r$) & $\Delta\omega_d$ & $\sim2\pi\times$ \SI{9.837}{\kilo\hertz} &  \\ \hline
Reduced E1 Matrix Element & $\braket{J \| er \| J'}$ & \SI{1.970(17)}{}\,$ea_0$ & \\ \hline
\makecell[l]{
  Rabi Frequency \\
\footnotesize $P=\SI{1}{\milli\watt}$, $D=\SI{1}{\milli\meter}$
} & $\Omega_0^{\ket{m_J=0} \text{\scalebox{0.7}[1]{$\to$}} \ket{m'_J=0}}$ & $2\pi\times$ \SI{20.16(17)}{\mega\hertz} &  \\ \hline
Isotope shift & $\nu({}^{87}\mathrm{Sr})-\nu({}^{88}\mathrm{Sr})$ & \SI{+7.29(30)}{\mega\hertz} & \cite{Courtillot2005, Baillard2007, Takano2017, Miyake2019} \\ \hline
\end{tabular}
\end{table}

\begin{table}[h!]
\renewcommand{\arraystretch}{1.15} 
\centering
\caption{$^3\mathrm{P}_1 \rightarrow\,^3\mathrm{S}_1$ Transition (\SI{688}{\nano\meter}).}
\begin{tabular}{|l|c|c|c|}
\hline
\textbf{Parameter} & \textbf{Symbol} & \textbf{Value} & \textbf{Ref.} \\ \hline
Frequency & $\omega_0$ & $2\pi\times$ \SI{435.7316972(5)}{\tera\hertz} &\cite{Courtillot2005}  \\ \hline
Transition Energy & $\hbar\omega_0$ & \SI{1.802041505(2)}{\electronvolt} &  \\ \hline
Wavelength (vacuum) & $\lambda$ & \SI{688.0207704(8)}{\nano\meter} &  \\ \hline
Wavelength (air) & $\lambda_{\rm air}$ & \SI{687.837765(7)}{\nano\meter} &  \\ \hline
Wavenumber (vacuum) & $k_L/2\pi$ & \SI{14534.444933(17)}{\centi\meter^{-1}} & \\ \hline
Lifetime excited state & $\tau$ & \makecell{$13.89(21)\,\mathrm{ns}$ \\ $R_{\mathrm{B}}$: 1.9} & \cite{Brinkmann1969, Havey1977, Jonsson1984, Heinz2020} \\ \hline
Decay rate & $\Gamma$ & \SI{72.0(11)e6}{\second^{-1}} &  \\ \hline
Natural Line Width & $\Delta\nu_{\rm FWHM}$ & $2\pi\times$ \SI{11.46(17)}{\mega\hertz} & \\ \hline
Branching Ratio & $\beta$ & $0.3405(24)$ & \cite{UDportal} \\ \hline
Partial decay rate & $\Gamma_{\rm partial}$ & \SI{24.5(4)e6}{\second^{-1}} &  \\ \hline
Recoil velocity & $v_r$ & $\sim$ \SI{6.598}{\milli\meter/\second} &  \\ \hline
Recoil energy & $\omega_r$ & $\sim2\pi\times$ \SI{4.795}{\kilo\hertz} &  \\ \hline
Recoil temperature & $T_r$ & $\sim$ \SI{460}{\nano\kelvin} &  \\ \hline
Doppler shift ($v=v_r$) & $\Delta\omega_d$ & $\sim2\pi\times$ \SI{9.589}{\kilo\hertz} &  \\ \hline
Reduced E1 Matrix Element & $\braket{J \| er \| J'}$ & \SI{3.439(29)}{}\,$ea_0$ & \\ \hline
\makecell[l]{
  Rabi Frequency \\
\footnotesize $P=\SI{1}{\milli\watt}$, $D=\SI{1}{\milli\meter}$
} & $\Omega_1^{\ket{m_J=0} \text{\scalebox{0.7}[1]{$\to$}} \ket{m'_J=1}}$ & $2\pi\times$ \SI{24.88(21)}{\mega\hertz} &  \\ \hline
Isotope shift & $\nu({}^{87}\mathrm{Sr})-\nu({}^{88}\mathrm{Sr})$ & \SI{+7.29(30)}{\mega\hertz} & \cite{Courtillot2005, Miyake2019} \\ \hline
\end{tabular}
\end{table}

\begin{table}[h!]
\renewcommand{\arraystretch}{1.15} 
\centering
\caption{$^3\mathrm{P}_2 \rightarrow\,^3\mathrm{S}_1$ Repump Transition (\SI{707}{\nano\meter}).}
\begin{tabular}{|l|c|c|c|}
\hline
\textbf{Parameter} & \textbf{Symbol} & \textbf{Value} & \textbf{Ref.} \\ \hline
Frequency & $\omega_0$ & $2\pi\times$ \SI{423.91634(3)}{\tera\hertz} & \cite{Onishchenko2019} \\ \hline
Transition Energy & $\hbar\omega_0$ & \SI{1.75317713(12)}{\electronvolt} &  \\ \hline
Wavelength (vacuum) & $\lambda$ & \SI{707.19721(5)}{\nano\meter} &  \\ \hline
Wavelength (air) & $\lambda_{\rm air}$ & \SI{707.00923(5)}{\nano\meter} & \\ \hline
Wavenumber (vacuum) & $k_L/2\pi$ & \SI{14140.3272(10)}{\centi\meter^{-1}} & \\ \hline
Lifetime excited state & $\tau$ & \makecell{$13.89(21)\,\mathrm{ns}$ \\ $R_{\mathrm{B}}$: 1.9} & \cite{Brinkmann1969, Havey1977, Jonsson1984, Heinz2020} \\ \hline
Decay rate & $\Gamma$ & \SI{72.0(11)e6}{\second^{-1}} &  \\ \hline
Natural Line Width & $\Delta\nu_{\rm FWHM}$ & $2\pi\times$ \SI{11.46(17)}{\mega\hertz} & \\ \hline
Branching Ratio & $\beta$ & $0.5432(25)$ & \cite{UDportal} \\ \hline
Partial decay rate & $\Gamma_{\rm partial}$ & \SI{39.1(6)e6}{\second^{-1}} &  \\ \hline
Recoil velocity & $v_r$ & $\sim$ \SI{6.419}{\milli\meter/\second} &  \\ \hline
Recoil energy & $\omega_r$ & $\sim2\pi\times$ \SI{4.538}{\kilo\hertz} &  \\ \hline
Recoil temperature & $T_r$ & $\sim$ \SI{436}{\nano\kelvin} &  \\ \hline
Doppler shift ($v=v_r$) & $\Delta\omega_d$ & $\sim2\pi\times$ \SI{9.076}{\kilo\hertz} &  \\ \hline
Reduced E1 Matrix Element & $\braket{J \| er \| J'}$ & \SI{4.526(36)}{}\,$ea_0$ & \\ \hline
\makecell[l]{
  Rabi Frequency \\
\footnotesize $P=\SI{1}{\milli\watt}$, $D=\SI{1}{\milli\meter}$
} & $\Omega_0^{\ket{m_J=0} \text{\scalebox{0.7}[1]{$\to$}} \ket{m'_J=0}}$ & $2\pi\times$ \SI{29.29(23)}{\mega\hertz} &  \\ \hline
Isotope shift & $\nu({}^{87}\mathrm{Sr})-\nu({}^{88}\mathrm{Sr})$ & \SI{8.0(3)}{\mega\hertz} & \cite{Courtillot2005, Trautmann2023} \\ \hline
\end{tabular}

\label{tab:last_repump_data}
\end{table}

\begin{table}[h!]
\renewcommand{\arraystretch}{1.15} 
\centering
\caption{$^1\mathrm{S}_0 \rightarrow\,^3\mathrm{P}_0$ Clock Transition (\SI{698}{\nano\meter}).}
\begin{tabular}{|l|c|c|c|}
\hline
\textbf{Parameter} & \textbf{Symbol} & \textbf{Value} & \textbf{Ref.} \\ \hline
Frequency & $\omega_0$ & $2\pi\times$ \SI{429.228 066 418 007 008(82)}{\tera\hertz}  & \cite{BIPM2025_88Sr429THz} \\ \hline
Transition Energy & $\hbar\omega_0$ & \SI{1.7751446488980402(3)}{\electronvolt} &  \\ \hline
Wavelength (vacuum) & $\lambda$ & \SI{698.445 608 419 382 44(13)}{\nano\meter} & \cite{BIPM2025_88Sr429THz} \\ \hline
Wavelength (air) & $\lambda_{\rm air}$ & \SI{698.259 898(7)}{\nano\meter} &  \\ \hline
Wavenumber (vacuum) & $k_L/2\pi$ & \SI{14317.507160837482(3)}{\centi\meter^{-1}} &  \\ \hline
Recoil velocity & $v_r$ & $\sim$ \SI{6.499}{\milli\meter/\second} &  \\ \hline
Recoil energy & $\omega_r$ & $\sim2\pi\times$\SI{4.653}{\kilo\hertz} &  \\ \hline
Recoil temperature & $T_r$ & $\sim$ \SI{447}{\nano\kelvin} &  \\ \hline
Doppler shift ($v=v_r$) & $\Delta\omega_d$ & $\sim2\pi\times$\SI{9.305}{\kilo\hertz} &  \\\hline
\makecell[l]{
  Rabi Frequency \\
\footnotesize $P=\SI{1}{\milli\watt}$, $D=\SI{1}{\milli\meter}$, $B=\SI{500}{G}$
} & $\Omega_{0}$ & $2\pi\times$ \SI{161.2(2)}{\hertz} &  \\ \hline
Isotope shifts & 
\makecell{$\nu({}^{84}\mathrm{Sr})-\nu({}^{88}\mathrm{Sr})$\\
          $\nu({}^{86}\mathrm{Sr})-\nu({}^{88}\mathrm{Sr})$\\
          $\nu({}^{87}\mathrm{Sr})-\nu({}^{88}\mathrm{Sr})$} & 
\makecell{\SI{-349.656(10)}{\mega\hertz}\\
          \SI{-162.939(11)}{\mega\hertz}\\
          \SI{-62.188 134 004(10)}{\mega\hertz}}& 
\makecell{\cite{Miyake2019}\\
          \cite{Miyake2019}\\
          \cite{Courtillot2005, Baillard2007, Takano2017, Miyake2019}} \\ \hline
\end{tabular}
\label{tab:clock_transition_data}
\end{table}

\begin{table}[h!]
\centering
\caption{$^1\mathrm{S}_0 \rightarrow\,^3\mathrm{P}_2$ M2 Transition (\SI{671}{\nano\meter}).}   
\renewcommand{\arraystretch}{1.15} 
\begin{tabular}{|l|c|c|c|}
\hline
\textbf{Parameter} & \textbf{Symbol} & \textbf{Value} & \textbf{Ref.} \\ \hline
Frequency & $\omega_0$ & $2\pi\times$ \SI{446.647242704(2)}{\tera\hertz} & \cite{Trautmann2023} \\ \hline
Transition Energy & $\hbar\omega_0$ & \SI{1.847184573571(8)}{\electronvolt} & \\ \hline
Wavelength (vacuum) & $\lambda$ & \SI{671.206 333 179(3)}{\nano\meter} &  \\ \hline
Wavelength (air) & $\lambda_{\rm air}$ & \SI{671.027 689(7)}{\nano\meter} &  \\ \hline
Wavenumber (vacuum) & $k_L/2\pi$ & \SI{14898.54833853(7)}{\centi\meter^{-1}} &  \\ \hline
Lifetime excited state & $\tau$ & \SI{520(310)}{\second} & \cite{Yasuda2004} \\ \hline
Decay rate & $\Gamma$ & \SI{1.9(11)e-3}{\second^{-1}} & \\ \hline
Natural Line Width & $\Delta\nu_{\rm FWHM}$ & $2\pi\times$ \SI{0.3(2)}{\milli\hertz} & \\ \hline
Branching ratio & $\beta$ & \SI{0.08(5)}{} &  \\ \hline
Partial decay rate & $\Gamma_{\rm partial}$ & \SI{152(43)e-6}{\second^{-1}} & \cite{Kluesener2024} \\ \hline
Recoil velocity & $v_r$ & $\sim$ \SI{6.763}{\milli\meter/\second} &  \\ \hline
Recoil energy & $\omega_r$ & $\sim2\pi\times$\SI{5.038}{\kilo\hertz} &  \\ \hline
Recoil temperature & $T_r$ & $\sim$ \SI{484}{\nano\kelvin} &  \\ \hline
Doppler shift ($v=v_r$) & $\Delta\omega_d$ & $\sim2\pi\times$\SI{10.076}{\kilo\hertz} &  \\ \hline
M2 Matrix Element & $\braket{^1S_0\|M2\|\,^3P_2}$ & $22.6(4) \,\mu_{\rm B}$ & \cite{Kluesener2024} \\ \hline
\makecell[l]{
  Rabi Frequency \\
\footnotesize $P\approx\SI{30}{\milli\watt}$, $D=\SI{1}{\milli\meter}$, $\theta=\SI{14(2)}{\degree}$
} & $\Omega_{\pm1}^{\ket{m_J=0} \text{\scalebox{0.7}[1]{$\to$}} \ket{m'_J=0}}$ & $2\pi\times$ \SI{189(1)}{\hertz} & \cite{Kluesener2024} \\ \hline
Isotope shift & $\nu({}^{87}\mathrm{Sr})-\nu({}^{88}\mathrm{Sr})$ & \SI{-62.91(4)}{\mega\hertz} & \cite{Trautmann2023} \\ \hline
\end{tabular}
\label{tab:m2_transition_data}
\end{table}

\begin{figure}[h!]
  \centering
  \includegraphics[width=1\textwidth]{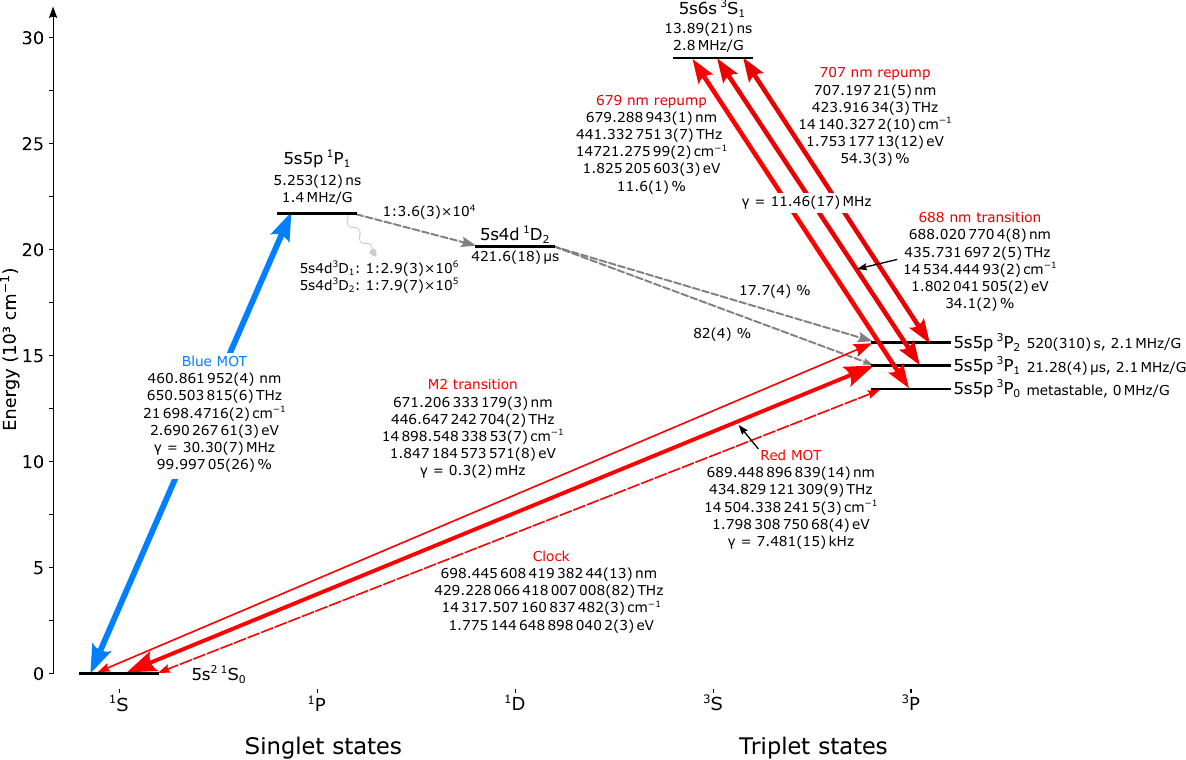}
  \caption{Data of the transitions discussed in this reference. Tables~\ref{tab:blue_transition_data}-\ref{tab:m2_transition_data} list the frequencies, wavelengths, energies, linewidths, and branching ratios of the transitions as well as the lifetime of the states with the corresponding references. The Zeeman splittings between adjacent magnetic sublevels can be found in Tab.~\ref{tab:zeeman_coeffs}. Note that for clarity the $^3\mathrm{P}_0$ state has been shifted downward and the $^3\mathrm{P}_2$ state upward relative to their true energies, the correct level energies appear in Fig.~\ref{fig:SrI-levels}.}
  \label{fig:Sr_I_energy_levels_reference}
\end{figure}

\begin{figure}[h!]
  \centering
  \includegraphics[width=1\textwidth]{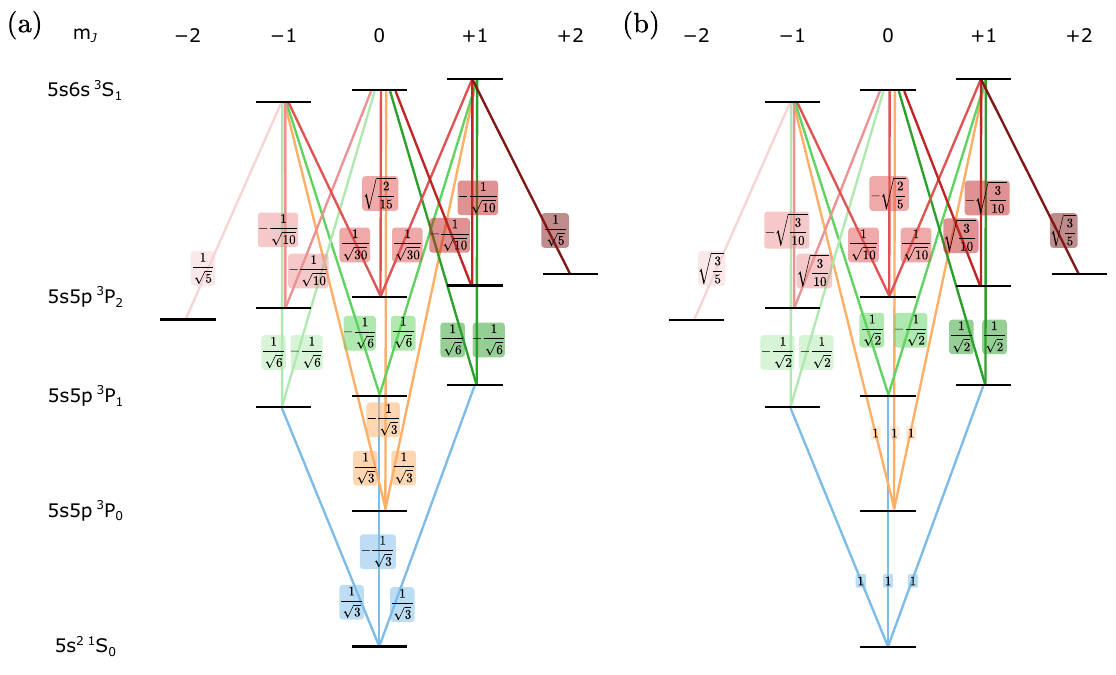}
  \caption{Polarization‐dependent coupling factors for the transition between various Zeeman levels.  
  (a) Wigner factors $w_q^{g\to e}$ as defined in Eq.~\ref{eq:wigner_factor_def}.  
  (b) The corresponding Clebsch–Gordan coefficients $\langle J_g\,m_{J_g};1\,q \mid J_e\,m_{J_e}\rangle$,
  given in Eq.~\ref{eq:clebsch_gordan}, for the same set of sublevels and polarizations. Note that a similar figure can be found in Ref.~\cite{Tang2022}.}
  \label{fig:cg_wigner}
\end{figure}

\clearpage

\section{Acknowledgments}

We thank Maximilian Ammenwerth, Robin Eberhard, Neven {\v S}anti\'c, Eric Yilun Song, Yoshio Torii, and Tristan Valenzuela-Salazar for stimulating discussions, comments, corrections, and suggestions. S.L.K. acknowledges support from the Independent Research Fund Denmark.

\setcounter{secnumdepth}{0}
\section{Release notes}
\noindent\textbf{v3}
\begin{itemize}
    \item Updated the recommended \(^1\mathrm{S}_{0}\!\rightarrow{}^1\mathrm{P}_{1}\) transition frequency in Tab.~\ref{tab:blue_transition_data} and Fig.~\ref{fig:Sr_I_energy_levels_reference} by combining a recent high-precision measurement with the historical value previously used in this reference.
    \item Corrected the \(^1\mathrm{P}_{1}\) lifetime data, revised the discussion of the consistency among the available measurements, and updated the recommended lifetime values in Tab.~\ref{tab:blue_transition_data} and Fig.~\ref{fig:Sr_I_energy_levels_reference}.
    \item Incorporated recent direct measurements of the \(^{1}\mathrm{P}_1\rightarrow{}^{1}\mathrm{D}_2\) decay rate and the \(^{1}\mathrm{D}_2\rightarrow{}^{3}\mathrm{P}_{1,2}\) branching fractions, and updated the corresponding values in the
    text, Tab.~\ref{tab:blue_transition_data}, and Fig.~\ref{fig:Sr_I_energy_levels_reference}.
    \item Updated the \(^{1}\mathrm{D}_2\) lifetime using a weighted mean of the available experimental measurements.
    \item Included the measured \(^{1}\mathrm{P}_1\rightarrow{}^{3}\mathrm{D}_{1,2}\) leakage channels in the calculation of the
    \(^{1}\mathrm{P}_1\rightarrow{}^{1}\mathrm{S}_0\) branching fraction.
    \item Updated the recommended $^1\mathrm{S}_0\rightarrow{}^3\mathrm{P}_0$ clock-transition frequency using the 2025 BIPM adjustment of standard frequencies.
    \item Updated the recommended \SI{813.4}{\nano\meter} clock-magic wavelength using only measurements that explicitly use \(^{88}\mathrm{Sr}\), excluding measurements performed in \(^{87}\mathrm{Sr}\) from the weighted mean. Added a discussion justifying the isotope-specific treatment of magic wavelengths.
    \item Added asterisks in Tab.~\ref{tab:special_wavelengths} to denote wavelength values measured entirely or partly using strontium isotopes other than $^{88}\mathrm{Sr}$.
    \item Added magic-wavelength measurements to Tab.~\ref{tab:special_wavelengths} for the ${}^3\mathrm{P}_2(m_J=2)\!-\!5\mathrm{s}4\mathrm{d}\,{}^3\mathrm{D}_3(m_J=3)$ transition and for the ${}^3\mathrm{P}_0\!-\!{}^3\mathrm{P}_2(m_J=0)$ fine-structure transition at \(\beta=90^\circ\).
    \item Added a reference for the \(^{86}\mathrm{Sr} - {}^{86}\mathrm{Sr}\) scattering length and updated the corresponding recommended value in Tab.~\ref{tab:sr_scattering_lengths}.
    \item Corrected the \(^{1}\mathrm{S}_{0}\!\rightarrow{}^{3}\mathrm{P}_{2}\) Rabi-frequency entry in Tab.~\ref{tab:m2_transition_data} by updating the probe power and correcting the Rabi-frequency unit.
    \item Replaced preprint citations with journal-published references, where available.
\end{itemize}

\noindent\textbf{v2}
\begin{itemize}
    \item Added a discussion of the recently demonstrated \SI{448}{\nano\meter} repumping scheme, including the reported improvement in MOT atom number.
    \item Added a polarization-dependent triple-magic condition near \SI{813.4}{\nano\meter} for the ${}^1\mathrm{S}_0$, ${}^3\mathrm{P}_0$, and ${}^3\mathrm{P}_1\,(m_J=0)$ states, together with a new entry in Tab.~\ref{tab:special_wavelengths}.
    \item Corrected the electronic-state label $5\mathrm{s}6\mathrm{s}\,{}^3\mathrm{S}_1$ in Fig.~\ref{fig:Sr_I_energy_levels_reference}.
    \item Corrected several typographical errors and made minor editorial, formatting, acknowledgment, and affiliation updates.
\end{itemize}

\noindent\textbf{v1}

\noindent Original release.

\printbibliography

\end{document}